\documentclass[twocolumn,nofootinbib]{revtex4-1}
\usepackage{graphicx,url,color,rotating,amssymb,amsmath,hyperref,bm}
\everymath{\displaystyle}

\newcommand{\mnras}{MNRAS}

\begin{document}

\title{Principal components of dark energy with SNLS supernovae:\\ the effects of systematic errors}

\author{Eduardo J. Ruiz}
\email{ejruiz@umich.edu}

\author{Daniel L. Shafer}
\email{dlshafer@umich.edu}

\author{Dragan Huterer}
\email{huterer@umich.edu}
\affiliation{Department of Physics, University of Michigan, 450 Church St, Ann Arbor, MI 48109-1040}

\author{Alexander Conley}
\email{alexander.conley@colorado.edu}
\affiliation{Center for Astrophysics and Space Astronomy, University of Colorado, 389 UCB, Boulder, CO 80309-0389}

\begin{abstract}
We study the effects of current systematic errors in Type Ia supernova
(SN Ia) measurements on dark energy (DE) constraints using current data from the Supernova Legacy Survey (SNLS). We consider
how SN systematic errors affect constraints from combined SN Ia,
baryon acoustic oscillations (BAO), and cosmic microwave background
(CMB) data, given that SNe Ia still provide the strongest constraints
on DE but are arguably subject to more significant systematics than
the latter two probes. We focus our attention on the temporal
evolution of DE described in terms of principal components (PCs) of
the equation of state, though we examine a few of the more common,
simpler parametrizations as well. We find that the SN Ia systematics
degrade the total generalized figure of merit (FoM), which
characterizes constraints in multi-dimensional DE parameter space, by
a factor of two to three. Nevertheless, overall constraints obtained
on more than five PCs are very good even with current data and
systematics. We further show that current constraints are robust to
allowing for the finite detection significance of the BAO feature in
galaxy surveys.
\end{abstract}

\maketitle

\section{Introduction}\label{sec:intro}

Since the discovery of the accelerating universe in the late 1990s
\cite{Riess_1998,Perlmutter_1999}, a tremendous amount of effort has been
devoted to improving measurements of dark energy (DE) parameters.  As
constraints on these parameters improved, controlling the systematic
errors in measurements became critical for continued progress. The systematics
come in many flavors, including a multitude of instrumental effects and astrophysical effects.

Type Ia supernovae (SNe Ia) were used to discover DE and still provide the
best constraints on DE. The advantage of SNe Ia relative to other cosmological
probes is that {\it every} SN provides a distance measurement and therefore
some information about DE. More recently, SN Ia observations have been joined
by measurements of baryon acoustic oscillations (BAO), which provide
exceedingly accurate measurements of the angular diameter distance in redshift
bins. Cosmic microwave background (CMB) anisotropies come mostly from high
redshift and are thus not particularly effective in probing DE, but they do
provide one measurement of the angular diameter distance to redshift $z \simeq
1100$ very accurately. Galaxy clusters also constrain DE usefully, while weak
gravitational lensing is expected to become one of the most effective probes
of DE in the near future. For recent comprehensive reviews of DE probes, see
\cite{FriTurHut,Weinberg:2012es}.

In this work, we are interested in studying the effect of SN Ia systematics on
DE constraints by including the {\it covariance} of measurements between
different SNe. The covariance includes primarily systematic errors, and for
the first time it has been quantified in depth by
\citet{Conley2011}. Including the effects of the systematic errors,
represented by nonzero covariance, weakens the overall constraints on model
parameters. Here we wish to explore the effect of systematic errors for
general models of DE described by a number of principal components (PCs) of
the equation of state, though we first consider these effects for simpler,
more commonly used descriptions of the DE sector. We choose to combine the SN
Ia data with BAO and CMB measurements and estimate the effects of {\it
  current} systematic errors in SN Ia observations. We then proceed to study
another systematic concern that is particularly relevant for BAO: whether the
finite significance of the detection of the BAO feature in various surveys,
when taken into account, weakens the constraints imposed on DE parameters.

While we closely follow the accounting for the SN Ia systematics from
\citet{Conley2011}, we note that several other analyses have considered the
effect of SN systematics. However, most of these analyses only studied the
effects of the systematic errors on the constant equation of state
(e.g.\ \cite{WoodVasey_2007,Constitution,SDSS_SN,Conley2011}) or included the
additional parameter $w_a$ to describe the variation of the equation of state
with time (e.g.\ \cite{Sullivan2011}). Notable exceptions are studies by
\citet{Davetal07} and \citet{Rubin_SCP}, which considered a number of specific
DE models with non-standard behavior, and \citet{Amanullah_Union2} and
\citet{Suzuki_SCP}, which parametrized the DE density in several redshift
bins. Here our goal is to go beyond any specific models and study the effects
of systematic errors in current data on DE constraints in the greatest
generality possible. While a truly model-independent description of the DE
sector is of course impossible, a description of the expansion history in
terms of 10 or so parameters -- which we adopt in this paper -- comes
close\footnote{We do not, however, consider allowing departures from general
  relativity; doing so would further generalize the treatment.}. In this
sense, our paper complements the recent investigations by Mortonson et
al.\ \cite{Mort_current,MHH_FoM} (see also \cite{Huterer_Cooray,Wang_Tegmark_2005,Zunckel_Trotta,Zhao_Huterer_Zhang,Hojjati:2009ab,Ishida:2010nk,Shafieloo:2012ht,Seikel:2012uu,Zhao:2012aw}),
which studied constraints on very general descriptions of DE using (a slightly
different set of) current data but without specific study of the effects of
systematic errors.

The paper is organized as follows. In Sec.~\ref{sec:data}, we describe the SN
Ia, BAO, and CMB data (and for BAO and CMB, the distilled observable
quantities) that we use in our analysis. In Sec.~\ref{sec:results}, we discuss
useful parametrizations of DE and compare constraints on the DE parameters
with and without systematic errors included in the analysis. In
Sec.~\ref{sec:BAO_detect}, we investigate the effects of the finite detection
significance of the BAO feature in galaxy surveys on the cosmological
parameter constraints. In Sec.~\ref{sec:conclude}, we summarize our
conclusions.

\section{Data Sets Used}\label{sec:data}

We begin by describing the data sets used in this analysis. We have
used three probes of DE: SNe Ia, BAO and CMB anisotropies.

\subsection{SN Ia Data and Covariance}\label{sec:data_SN}

Although SNe Ia are not, of course, perfect standard candles, it has long been
known that there exist useful correlations between the peak apparent magnitude
of a SN Ia and the {\it stretch}, or broadness, of its light curve (simply
put, broader is brighter). The peak apparent magnitude is also correlated with
the color of the light curve (bluer is brighter). We therefore model the
apparent magnitude of a SN Ia with the equation \cite{Guy:2007dv}
\begin{equation}
m_{\mathrm{mod}} = 5 \log_{10} \left( \frac{H_0}{c} \, d_L \right) - \alpha_s (s - 1) + \beta_c \; \mathcal{C} + \mathcal{M},
\label{eq:mag}
\end{equation}
where $d_L$ is the luminosity distance, $\alpha_s$ is a nuisance parameter
associated with the measured stretch $s$ of a SN Ia light curve, and $\beta_c$
is a nuisance parameter associated with the measured color $\mathcal{C}$ of
the light curve. The absolute magnitude of a SN Ia is contained within the
constant magnitude offset $\mathcal{M}$, which is considered yet another
nuisance parameter\footnote{Throughout the analyses in this paper, we actually
  marginalize analytically over a model with {\it two} distinct $\mathcal{M}$
  values, where a mass cut of the host galaxy dictates which $\mathcal{M}$
  value applies (here we use a mass cut of $10^{10} M_\odot$). This is meant
  to correct for host galaxy properties and is empirical in nature (see text
  and Appendix C of \cite{Conley2011}). For simplicity, we suppress mention of
  the second $\mathcal{M}$ parameter.}.

Recent work has concentrated on estimating correlations between measurements of
individual SN Ia magnitudes. A complete covariance matrix for SNe Ia includes all identified sources of systematic error in addition to the intrinsic scatter and other sources of statistical error. The $\chi^2$ statistic is then given by
\begin{equation}
\chi^2 = \Delta \mathbf{m}^T \mathbf{C}^{-1} \Delta \mathbf{m},
\label{eq:covchi2}
\end{equation}
where $\Delta \mathbf{m} = \mathbf{m}_{\mathrm{obs}} -
\mathbf{m}_{\mathrm{mod}}(\mathbf{p})$ is the vector of magnitude differences
between the observed magnitudes of $N$ SNe Ia $\mathbf{m}_{\mathrm{obs}}$ and
the theoretical prediction that depends on the set of cosmological parameters
$\mathbf{p}$, $\mathbf{m}_{\mathrm{mod}}(\mathbf{p})$. Here $\mathbf{C}$ is
the $N \times N$ covariance matrix between the SNe. Given a value for
$\chi^2$, we assume that the likelihood of a set of cosmological parameters is
Gaussian, so that $\mathcal{L}(\mathbf{p}) \propto e^{-\chi^2/2}$.

\begin{figure}[t]
\includegraphics[width=.48\textwidth]{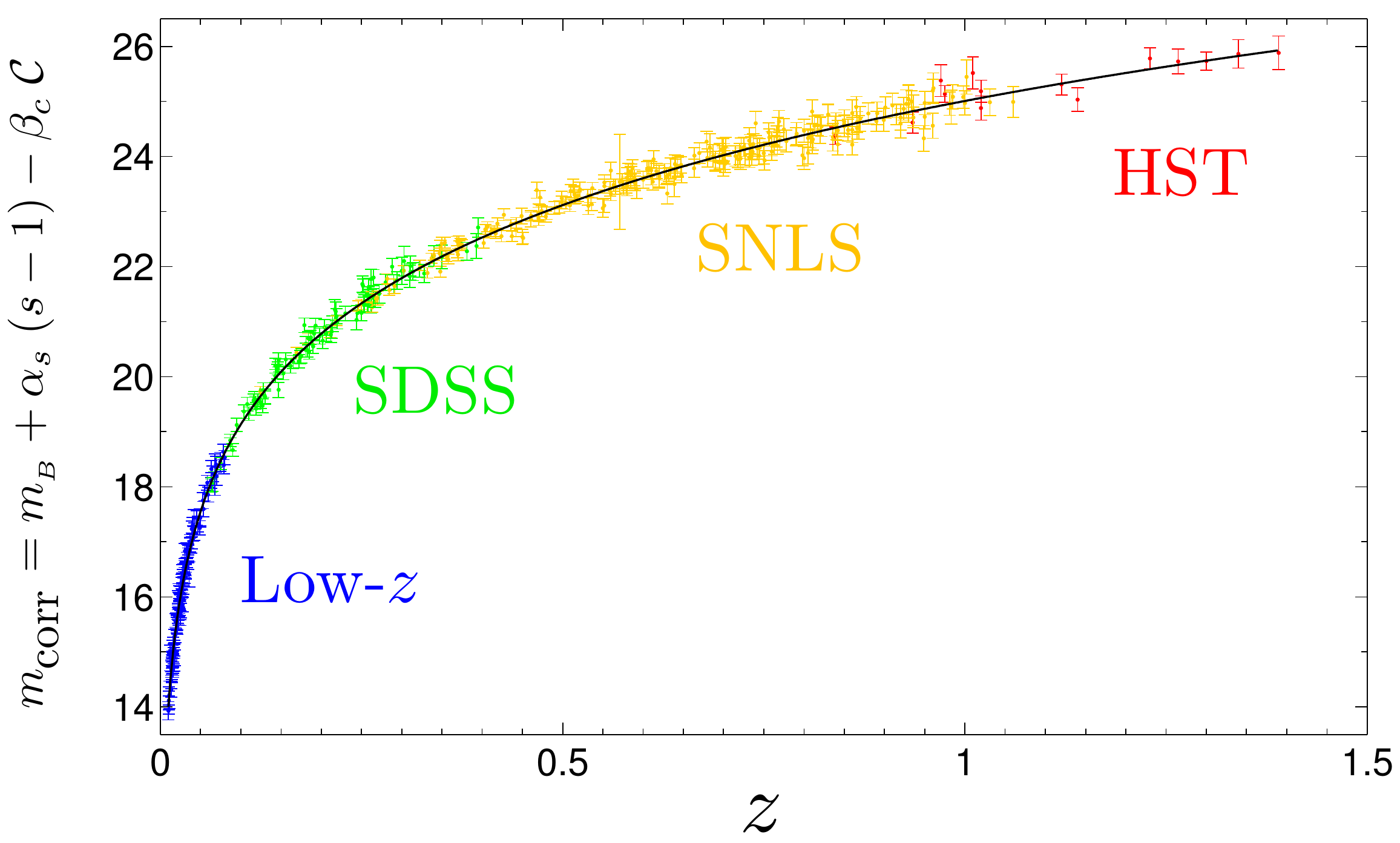}
\caption{Hubble diagram for the compilation of all SN Ia data used in this paper, labeling SNe from each survey separately and showing the (diagonal-only) magnitude uncertainties. The solid black line represents the best fit to the data.}
\label{fig:hubble}
\end{figure}

Recently \citet{Conley2011} determined covariances between SN Ia measurements
from the Supernova Legacy Survey (SNLS). The SN compilation and covariance
matrix that resulted from this work will be used in this analysis. The SNLS
compilation consists of 472 SNe Ia, approximately one half of which were
detected in the SNLS, while the rest originated from one of three other
sources. These four main sources are summarized in Table~\ref{tab:sndata} and
illustrated in the Hubble diagram of Fig.~\ref{fig:hubble}. The low-redshift
(Low-$z$) SNe actually come from a variety of samples as discussed in
\citet{Conley2011}.

\begin{table}[t]
\begin{center}
\setlength{\tabcolsep}{1em}
\begin{tabular}{|| c | c | c ||}
\hline \hline
\rule[-3mm]{0mm}{8mm} Source & $N_{\mathrm{SN}}$ & Range in $z$ \\ \hline \hline
\rule[-3mm]{0mm}{8mm} Low-$z$ & 123 &  0.01 - 0.1 \\ \hline
\rule[-3mm]{0mm}{8mm} SDSS & 93 & 0.06 - 0.4 \\ \hline
\rule[-3mm]{0mm}{8mm} SNLS & 242 & 0.08 - 1.05 \\ \hline
\rule[-3mm]{0mm}{8mm} HST & 14 & 0.7 - 1.4 \\ \hline
\end{tabular}
\end{center}
\caption{Summary of SN Ia observations included in this analysis, showing the number of SNe included from each survey and the approximate redshift ranges.}
\label{tab:sndata}
\end{table}

The complete covariance matrix from \cite{Conley2011} can be written
most usefully as the sum of two separate parts, a diagonal part
consisting of typical statistical errors and an off-diagonal
systematic part. The off-diagonal piece includes some correlated
errors which are considered statistical in \cite{Conley2011} (since
they can be reduced by including more observations), but here we
disregard the distinction and group these errors with the actual
systematic errors, which also lead to off-diagonal covariance
elements. This simplification is reasonable because the correlated
statistical errors are small compared to the (correlated) systematic errors. The diagonal (statistical-only)
part of the covariance matrix can be expressed as
\begin{align}
D^\mathrm{stat}_{ii} &= \sigma_{m_B,i}^2 + \alpha_s^2 \, \sigma_{s,i}^2 + \beta_c^2 \, \sigma_{\mathcal{C},i}^2 + \sigma_{\mathrm{int}}^2 \nonumber \\[0.2cm]
&+ \left(\frac{5(1+z_i)}{z_i(1+z_i/2) \log 10}\right)^2 \sigma_{z,i}^2 + \sigma^2_\mathrm{lensing} \label{eq:dstat} \\[0.2cm]
&+ \sigma^2_\mathrm{host \; correction} + D^{m_B s \,\mathcal{C}}_{ii}(\alpha_s,\beta_c) \nonumber
\end{align}
In the above, $\sigma_{m_B,i}\,$, $\sigma_{s,i}\,$, $\sigma_{\mathcal{C},i}\,$, and
$\sigma_{z,i}$ are the statistical uncertainties of the measured magnitude,
stretch, color, and redshift, respectively, of the $i^{\mathrm{th}}$ SN. The
$z$ term translates the error in redshift into error in magnitude. To include actual intrinsic scatter of SNe Ia and allow for any mis-estimates of photometric uncertainties, the quantity $\sigma_{\mathrm{int}}$ is included, with a different value allowed for each sample. Also included are statistical uncertainties due to gravitational lensing and uncertainty in the host galaxy correction.

The contribution to the (diagonal part of the) covariance matrix $D^{m_B s \,\mathcal{C}}_{ii}(\alpha_s,\beta_c)$ represents a combination of the covariance terms between magnitude, stretch, and color for the $i^{\mathrm{th}}$ SN. It is given by
\begin{align}
D^{m_B s \,\mathcal{C}}_{ii}(\alpha_s,\beta_c) = 2 \alpha_s D^{m_B \, s}_{ii} &- 2 \beta_c D^{m_B \,\mathcal{C}}_{ii} \\[0.2cm]
&- 2 \alpha_s \beta_c D^{s \,\mathcal{C}}_{ii}. \nonumber
\end{align}
Note that even the diagonal covariance matrix is a function of $\alpha_s$ and $\beta_c$, meaning that a proper analysis involves varying the errors (recomputing the covariance matrix) any time $\alpha_s$ and $\beta_c$ are changed.

\begin{figure*}[t]
\includegraphics[width=0.45\textwidth]{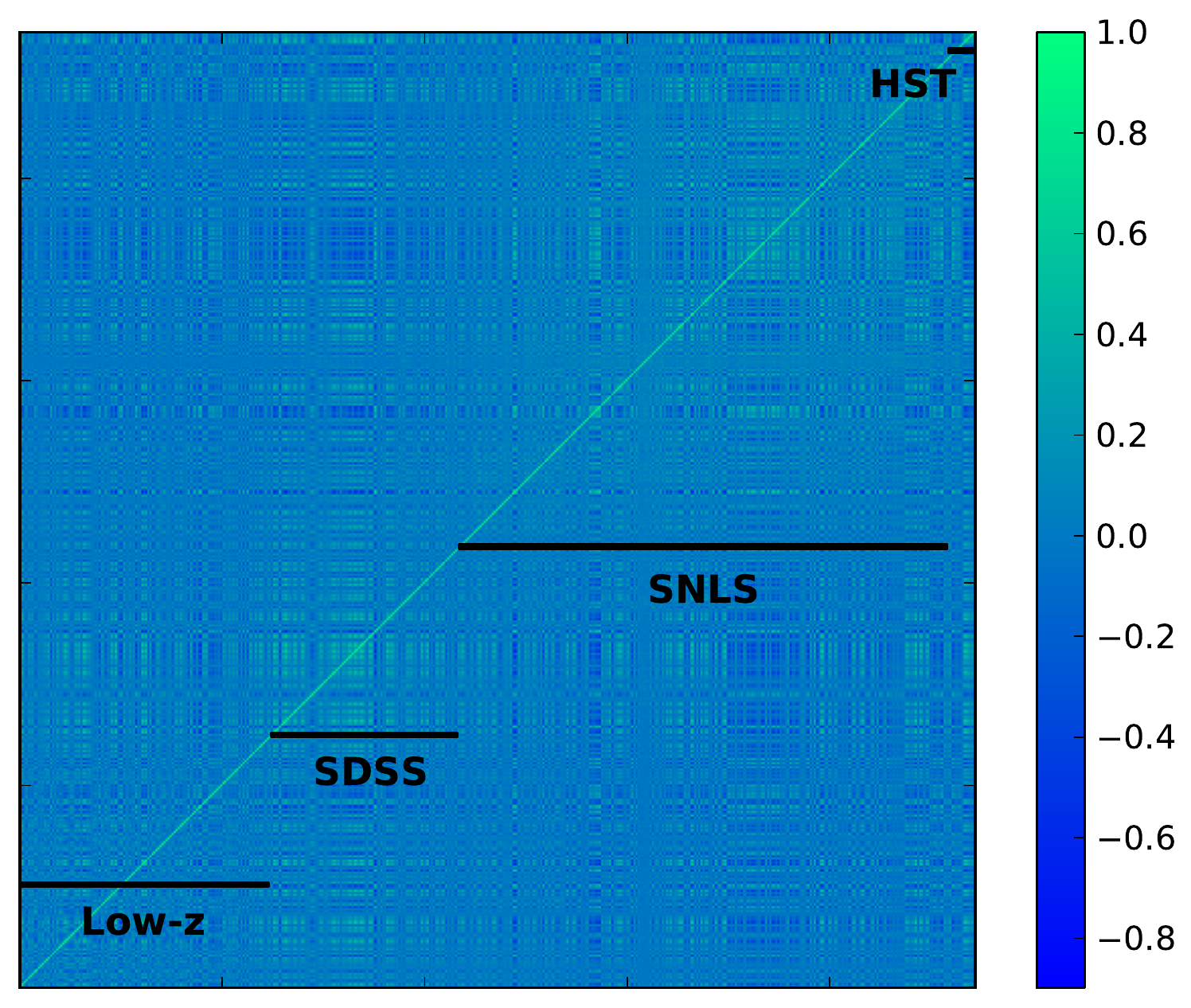}\hspace{0.3cm}
\includegraphics[width=0.45\textwidth]{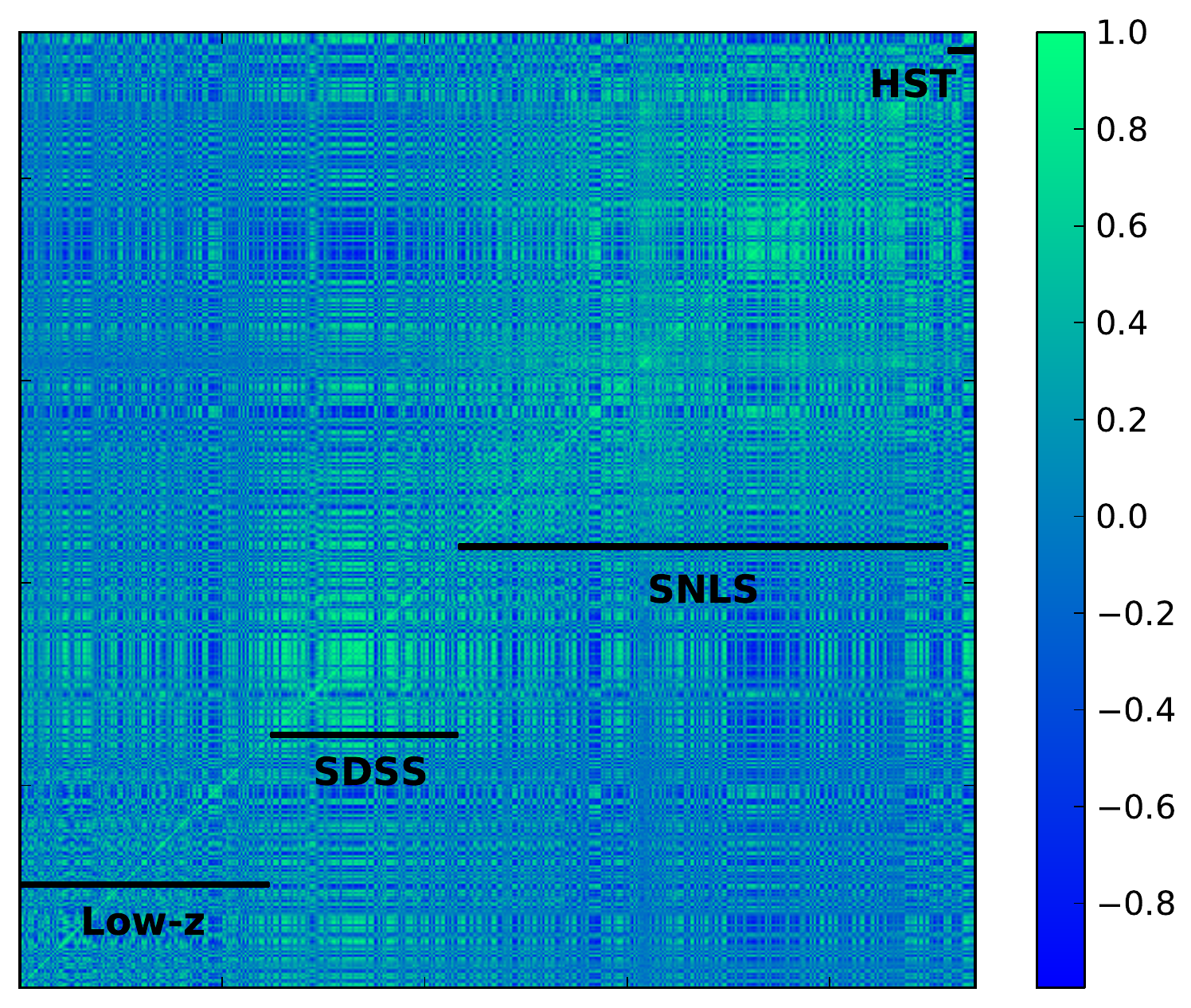}
\caption{Left panel: correlation matrix obtained from the complete covariance
  matrix $\mathbf{C}^\mathrm{full}$, sorted first by survey and then by
  redshift within each survey. Right panel: same, but using only the
  systematic covariance matrix $\mathbf{C}^\mathrm{sys}$. In both cases we
  assume $\alpha_s = 1.43$ and $\beta_c = 3.26$, the best-fit values for the flat
  $w = \mathrm{const}$ model. The right panel is similar to Fig.~12 from
  \cite{Conley2011}, but we repeat it here and show the full covariance (left
  panel) for completeness.}
\label{fig:covmatrixplot}
\end{figure*}

A similar equation can be used to construct the off-diagonal
systematic covariance matrix, where different systematic terms are
combined to produce submatrices which are then added together with
specified values for $\alpha_s$ and $\beta_c$, as above. The elements
of the systematic covariance matrix (see \cite{Conley2011} for
more details) are calculated using the equation
\begin{equation}
C^{\mathrm{sys}}_{i j} = \sum_{k = 1}^K
\left( \frac{ \partial m_{\mathrm{mod}\, i} }{ \partial S_k } \right)
\left( \frac{ \partial m_{\mathrm{mod}\, j} }{ \partial S_k } \right)
\left( \Delta S_k \right)^2,
\end{equation}
where the sum is over the $K$ systematics $S_k$, $\Delta S_k$ is the size of
each term (for example, the uncertainty in the zero point), and
$m_{\mathrm{mod}}$ is defined in Eq.~\eqref{eq:mag}. Then the full covariance
matrix is simply given by
\begin{equation}
\mathbf{C}^\mathrm{full} = \mathbf{D}^\mathrm{stat} + \mathbf{C}^\mathrm{sys}.
\label{eq:Cov}
\end{equation}
A plot of the full covariance matrix (constructed using flat $w = \mathrm{const}$ model best-fit
values $\alpha_s = 1.43$ and $\beta_c = 3.26$) is shown in Fig.~\ref{fig:covmatrixplot}.

\subsection{BAO and CMB data}

To produce the combined constraints in this paper, we include information from
both BAO and the CMB in addition to the SN data. In each case, we choose for
simplicity distilled quantities which depend only on $\Omega_M$,
$\Omega_{\mathrm{DE}}$, $\Omega_K$, and a parametrized $w(z)$.

For BAO, we compare the theoretical prediction for the acoustic parameter $A(z)$
with the measured value, where we define (see \citet{Eisenstein})
\begin{equation}
A(z) \equiv \left[r^2(z)\frac{cz}{H(z)}\right]^{1/3} \frac{\sqrt{\Omega_M H_0^2}}{cz}.
\label{eq:Az}
\end{equation}
We combine recent measurements of $A(z)$ at different effective redshifts,
using data from the 6dF Galaxy Survey \cite{Beutler:2011hx}, the Sloan Digital
Sky Survey (SDSS) Data Release 7 (DR7) \cite{Percival:2009xn}, the WiggleZ
survey \cite{Blake:2012pj,Blake:2011en}, and the SDSS Baryon Oscillation
Spectroscopic Survey (BOSS) \cite{Sanchez:2012sg,Anderson:2012sa}. The
measured values are summarized in Table~\ref{tab:baovals}.

\begin{figure}[h]
\includegraphics[width=.45\textwidth]{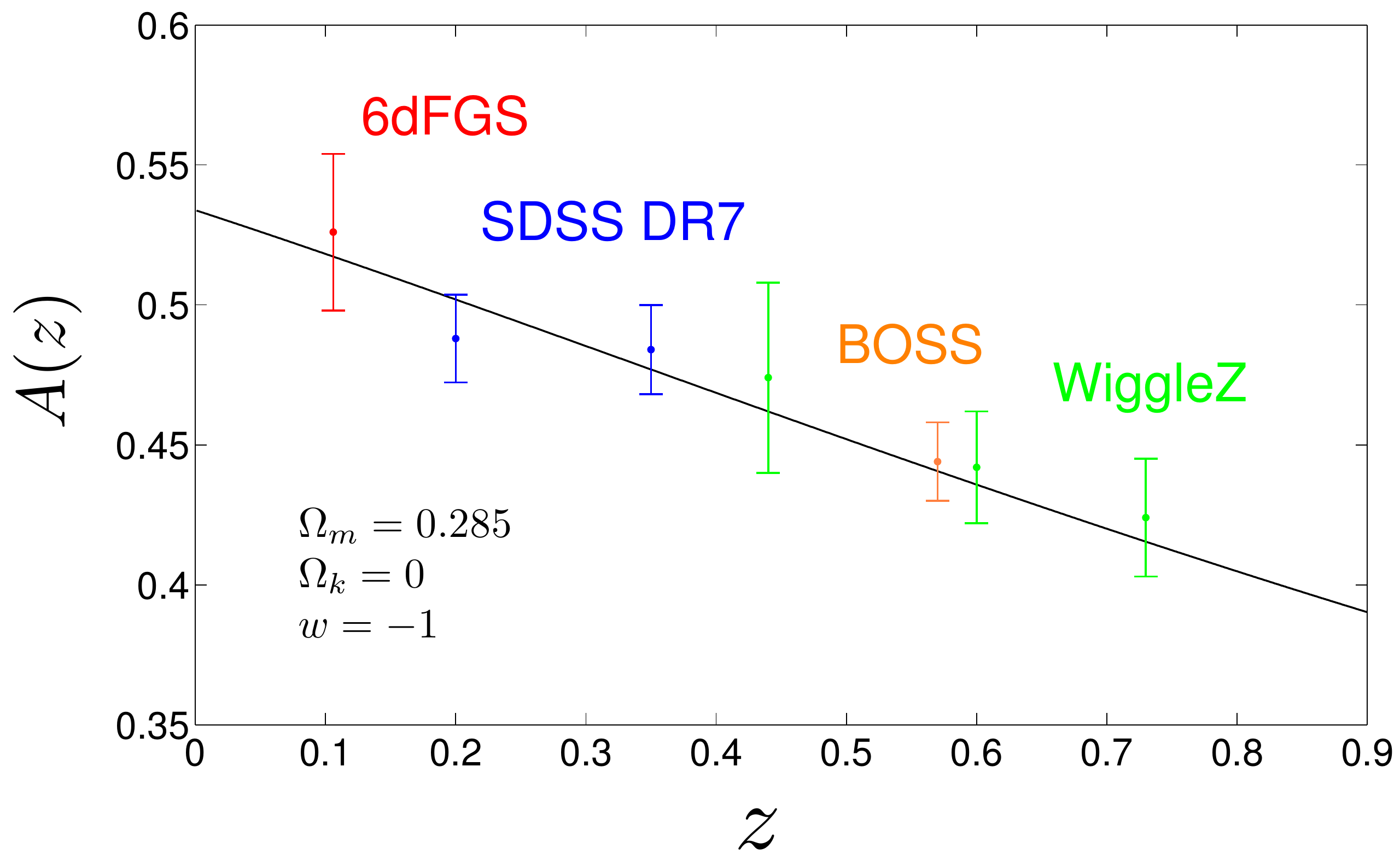}
\caption{Measured values of $A(z)$ and their (diagonal-only) uncertainties for each effective redshift. The black curve shows $A(z)$ for a model that fits the data points well, and the parameters for this model are given in the legend.}
\label{fig:Ameas}
\end{figure}

A plot of the measured values and their uncertainties superimposed on an
$A(z)$ curve (Fig.~\ref{fig:Ameas}) suggests that there is no significant
tension between the measurements. Note that the SDSS DR7 measurements at $z =
(0.2,0.35)$ are correlated with correlation coefficient 0.337. The WiggleZ
measurements are correlated with coefficient 0.369 for the pair $z =
(0.44,0.6)$ and coefficient 0.438 for $z = (0.6,0.73)$. Ignoring the
relatively small overlap in survey volume between SDSS DR7 and the BOSS
sample, we expect all other pairwise correlations to be zero. We compute
$\chi^2$ in the usual way for correlated measurements, as in Eq.~\eqref{eq:covchi2}.

\begin{table}[t]
\begin{center}
\setlength{\tabcolsep}{1em}
\begin{tabular}{|| c | c | c ||}
\hline \hline
\rule[-3mm]{0mm}{8mm} Sample & $z_{\mathrm{eff}}$ & $A_0(z_{\mathrm{eff}})$  \\ \hline \hline
\rule[-3mm]{0mm}{8mm} 6dFGS & $0.106$ & $0.526 \pm 0.028$  \\ \hline
\rule[-3mm]{0mm}{8mm} SDSS DR7 & $0.20$ & $0.488 \pm 0.016$ \\ \hline
\rule[-3mm]{0mm}{8mm} SDSS DR7 & $0.35$ & $0.484 \pm 0.016$ \\ \hline
\rule[-3mm]{0mm}{8mm} WiggleZ & $0.44$ & $0.474 \pm 0.034$ \\ \hline
\rule[-3mm]{0mm}{8mm} BOSS & $0.57$ & $0.444 \pm 0.014$ \\ \hline
\rule[-3mm]{0mm}{8mm} WiggleZ & $0.60$ & $0.442 \pm 0.020$ \\ \hline
\rule[-3mm]{0mm}{8mm} WiggleZ & $0.73$ & $0.424 \pm 0.021$ \\ \hline
\end{tabular}
\end{center}
\caption{Summary of measurements of distilled BAO parameter $A(z)$. We show
  the survey from which the measurement comes, the effective redshift of the
  survey (or its subsample), and the measured value $A_0$.}
\label{tab:baovals}
\end{table}

Nearly all of the sensitivity of the CMB to DE comes from the measurement of
an angle at which the sound horizon at $z \approx 1100$ is observed
(e.g.\ \cite{FriHutLinTur}). This measurement in turn determines the angular
diameter distance to recombination with the physical matter quantity,
$\Omega_M h^2$, essentially fixed. The latter quantity is popularly known as
the CMB shift parameter $R$ and is defined as
\begin{equation}
R \equiv \frac{\sqrt{\Omega_M H_0^2}}{c} \; r(z_\ast),
\label{eq:R}
\end{equation}
where $z_\ast = 1091.3$ is the redshift of decoupling as measured by WMAP7
\cite{wmap7} and $r(z)$ is the comoving distance to redshift $z$. We take the
measured value of $R$ to be the value determined by WMAP7, $R_0 = 1.725 \, \pm
\, 0.0184$ \cite{wmap7}. We compute $\chi^2$ in the usual way, comparing this
measured value of $R$ with the theoretical prediction.

Calculating the combined SN, BAO, and CMB likelihood is now a simple
task. We define $\mathcal{L}_{\mathrm{comb}} \propto
e^{-\chi^2_\mathrm{tot}/2}$, where $\chi^2_\mathrm{tot} =
\chi^2_\mathrm{SN} + \chi^2_\mathrm{BAO} + \chi^2_\mathrm{CMB}$.

\subsection{Parameter constraint methodology}

We use two alternate codes to produce our constraints. For the basic
constraints, including the constant equation of state of DE or the
$(w_0, w_a)$ description, we use a brute-force search which computes
likelihoods over a grid of values of $\sim \,$5 parameters (listed
below).

Alternatively, we developed a new Markov Chain Monte Carlo (MCMC;
e.g.\ see~\cite{Christensen:2001gj,Dunetal05}) code to determine DE parameter
constraints and figures of merit (FoMs) for the general ($\sim \,$13
parameters) PC description. The MCMC procedure is based on the
Metropolis-Hastings algorithm \cite{Metropolis:1953am,hastings}. From the
likelihood $\mathcal{L}(\mathbf{x}|\mathbf{\theta})$ of the data $\mathbf{x}$
given each proposed parameter set $\mathbf{\theta}$, Bayes' Theorem tells us
that the posterior probability distribution of the parameter set given the
data is
\begin{equation}
\mathcal{P}(\mathbf{\theta}|\mathbf{x}) =
\frac{\mathcal{L}(\mathbf{x}|\mathbf{\theta})\mathcal{P}(\mathbf{\theta})}{\int \mathcal{L}(\mathbf{x}|\mathbf{\theta})\mathcal{P}(\mathbf{\theta}) \, d\mathbf{\theta}},
\label{eq:bayes}
\end{equation}
where $\mathcal{P}(\mathbf{\theta})$ is the prior probability density. The
MCMC algorithm generates random draws from the posterior distribution. We test
convergence of the samples to a stationary distribution that approximates
$\mathcal{P}(\mathbf{\theta}|\mathbf{x})$ by applying a conservative
Gelman-Rubin criterion \cite{gelman/rubin} of $R-1 \lesssim 0.03$ across a
minimum of four chains for each model class. We use the {\tt getdist} routine
of the CosmoMC code \cite{Lewis:2002ah,cosmomc_url} to process the resulting
chains; {\tt getdist} bins the chains and then smoothes the binned
distribution of counts by convolution with a multidimensional Gaussian
kernel.

We verified that the two codes give results that are in excellent agreement in
several relevant cases, e.g.\ constraints in the $\Omega_M$--$w$ or
$w_0$--$w_a$ plane.

\section{Results: Effects Of The Systematics}\label{sec:results}

\subsection{Preliminaries}

Before beginning our discussion of systematics, we briefly consider the
vanilla $\Lambda \mathrm{CDM}$ cosmology, where $w = -1$. The cosmological parameters
describing the expansion rate are matter and cosmological constant densities
relative to critical, $\Omega_M$ and $\Omega_\Lambda$. Including the nuisance
parameters, the total parameter set is
\begin{equation}
p_i \in \{\Omega_M, \Omega_\Lambda, \mathcal{M}, \alpha_s, \beta_c \}.
\label{eq:params}
\end{equation}
We combine SN constraints with BAO and CMB constraints and marginalize over the other parameters to
map the likelihood of $\Omega_\Lambda$. We find a mean value
$\Omega_\Lambda = 0.724 \, \pm \, 0.0114$. This suggests that a universe
with zero (or negative) cosmological constant is ruled out at
approximately 64-$\sigma$! Amusingly, using the brute-force likelihood
search that includes the positive and negative values of
$\Omega_\Lambda$, we find that the combined data give a remarkably low
likelihood of zero or negative vacuum energy, even allowing for
nonzero curvature: $P(\Omega_\Lambda \leq 0) \sim 10^{-267}$. Of
course, in reality, the evidence for DE is not nearly this convincing,
since the likelihood in the space of cosmological observables is
certainly not expected to be Gaussian this far away from the peak and
thus would {\it not} be described by $\mathcal{L}_{\mathrm{comb}}
\propto e^{-\chi^2_\mathrm{tot}/2}$ (we discuss a related issue in
Sec.~\ref{sec:BAO_detect}). Nonetheless, it is impressive how strong
the evidence for DE is with current data.

We now discuss how one goes beyond $\Lambda \mathrm{CDM}$ cosmology by
parametrizing the DE equation of state.

Previous work on the effect of systematics, such as \cite{Conley2011},
considered the DE sector parametrized by its energy density relative to
critical, $\Omega_{\mathrm{DE}}$, and a constant equation of state $w$. Here,
we are particularly interested in extending the DE sector to allow for a
time-varying equation of state. We make two alternative choices in addition to
the constant equation of state so that the three parametrizations we consider
are:
\begin{enumerate}
\item Constant equation of state, $w = \mathrm{constant}$;
\item Equation of state described with $w_0$ and $w_a$ \cite{Linder_wa}, so that $w(a) = w_0 + w_a(1-a)$; 
\item Equation of state described by a finite number of principal components of $w(z)$ \cite{Huterer_Starkman}.
\end{enumerate}

We now describe in more detail the different parametrizations of DE that we
consider (constant $w$, $w_0$ and $w_a$, PCs) and then proceed to analyze the
effects of SN systematics on parameter constraints.

\subsection{Constant \texorpdfstring{$w$}{w}}

Assuming that DE can be described by an equation of state $w$ that is constant
in time, and assuming a flat universe, we calculate the SN-only likelihood in
the $\Omega_M$--$w$ plane. We marginalize over the usual nuisance parameters
$\mathcal{M}$, $\alpha_s$, and $\beta_c$.

\begin{figure}[t]
\includegraphics[width=.45\textwidth]{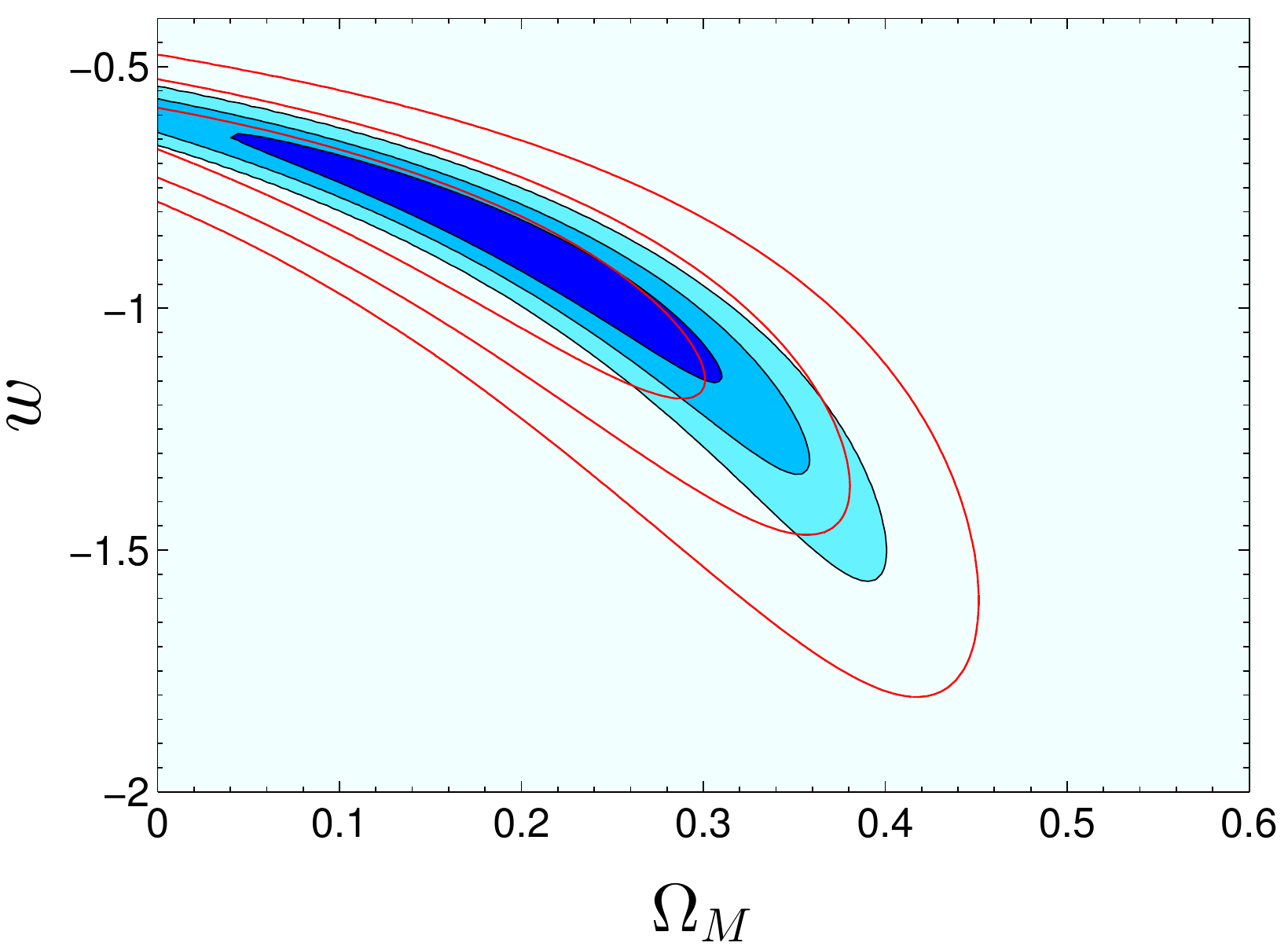}
\caption{68.3\%, 95.4\%, and 99.7\% likelihood constraints on $\Omega_M$ and
  $w$, assuming a constant value for $w$ and a flat universe. We use only SN
  data and marginalize over the nuisance parameters. We compare the case of
  diagonal statistical errors only (shaded blue) with the full systematic
  covariance matrix (red).}
\label{fig:Omw}
\end{figure}

The results for SN-only constraints on $\Omega_M$ and $w$ are shown in
Fig.~\ref{fig:Omw}, where we illustrate the effect of the systematics
by showing constraints from the full covariance matrix
$\mathbf{C}^\mathrm{full}$ on top of those which assume only the
diagonal statistical uncertainties $\mathbf{D}^\mathrm{stat}$. The
systematic uncertainties broaden the well-determined direction in the
$\Omega_M$-$w$ plane without elongating the poorly determined
direction much. Constraints in either parameter are not appreciably
shifted. The marginalized uncertainty for $w$ is $\sigma_w = 0.17$ for
diagonal errors only and $\sigma_w = 0.20$ when systematic errors are
included. Thus systematic errors increase the uncertainty by about 20\%.

\begin{figure}[t]
\includegraphics[width=.45\textwidth]{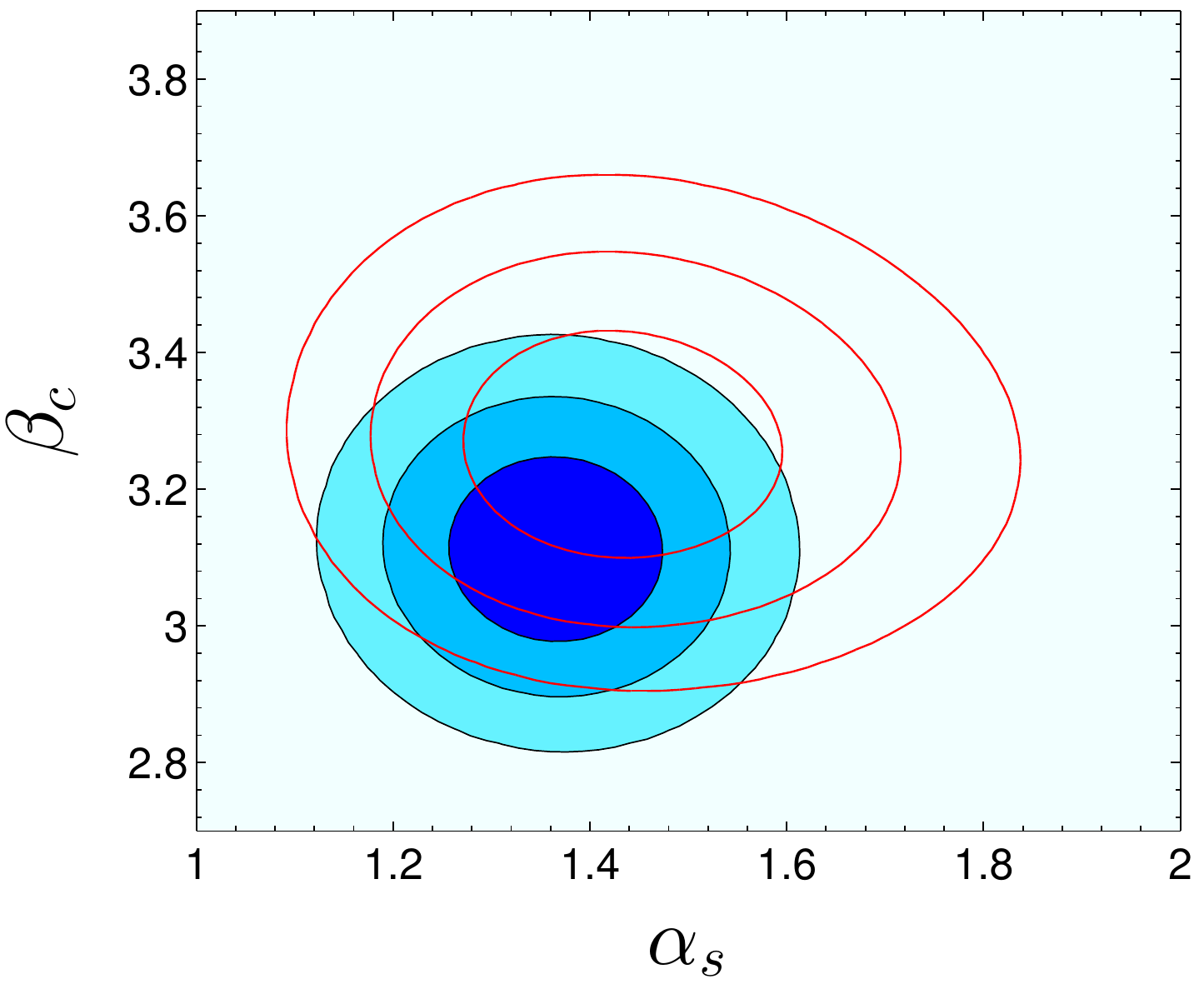}
\caption{68.3\%, 95.4\%, and 99.7\% likelihood constraints on $\alpha_s$ and
  $\beta_c$, assuming a constant value for $w$ and a flat universe. We use
  only SN data and marginalize over $\mathcal{M}$, $\Omega_M$, and $w$. We
  compare the case of diagonal statistical errors only (shaded blue) with the
  full systematic covariance matrix (red).}
\label{fig:ab}
\end{figure}

\begin{figure*}	
\includegraphics[width=\textwidth]{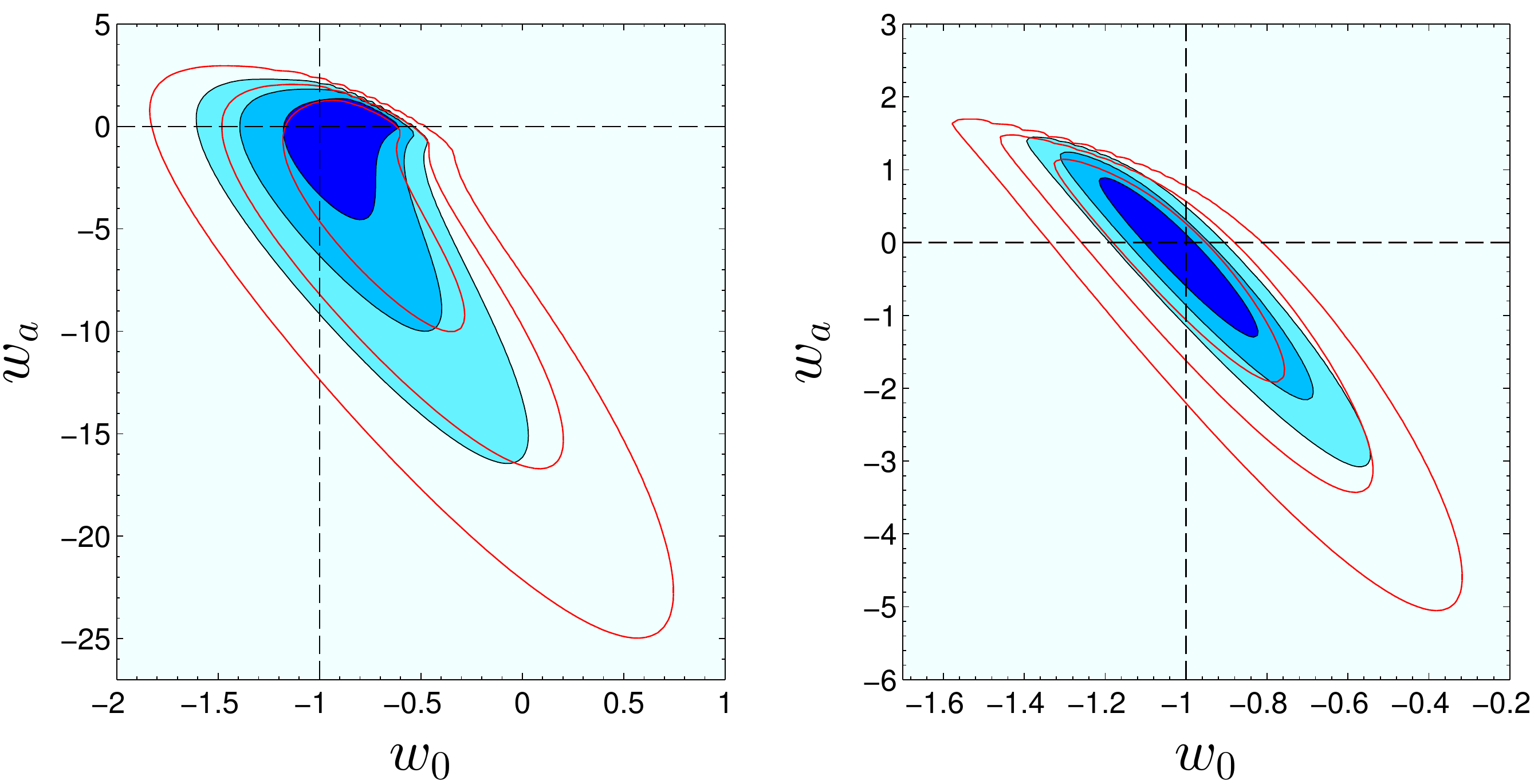}
\caption{68.3\%, 95.4\%, and 99.7\% likelihood constraints on $w_0$ and $w_a$
  in a flat universe, marginalized over $\Omega_M$ and the nuisance
  parameters. The left panel shows SN-only constraints, while the right panel
  shows combined SN+BAO+CMB constraints. The shaded blue contours represent
  constraints with only statistical SN errors assumed
  ($\mathbf{D}^\mathrm{stat}$), while the red contours represent the full
  systematic SN covariance matrix ($\mathbf{C}^\mathrm{full}$). Note that the
  $\Lambda \mathrm{CDM}$ model $(w_0,w_a) = (-1,0)$, represented by the black
  dashed lines, is fully consistent with the data.}
\label{fig:w0wa}
\end{figure*}

We also seek to understand how SN systematics influence the stretch and color
parameters $\alpha_s$ and $\beta_c$, not only because these correlations are
what make SNe Ia useful standard candles, but also because it is expected that
systematics could potentially affect these correlations. In Fig.~\ref{fig:ab},
we marginalize over $\mathcal{M}$, $\Omega_M$, and $w$ to show constraints on
the stretch and color coefficients $\alpha_s$ and $\beta_c$. Of particular
interest is the color coefficient $\beta_c$, which is broadly consistent with
values found previously; the systematic errors shift it slightly upwards and
increase errors in both parameters by a modest amount.

\subsection{\texorpdfstring{$w_0$ and $w_a$}{w0 and wa}}

We wish to understand the constraints on the redshift dependence of $w(z)$, so we
allow $w(z)$ to have the form \cite{Linder_wa,Chevallier_Polarski}
\begin{equation}
w(z) = w_0 + w_a \: z/(1+z).
\label{eq:wa}
\end{equation}
Constraints on $w_0$ and $w_a$ in a flat universe are shown in
Fig.~\ref{fig:w0wa}. The shaded blue contours represent constraints with only
statistical SN errors assumed ($\mathbf{D}^\mathrm{stat}$), while the red
contours include the full systematic SN covariance matrix
($\mathbf{C}^\mathrm{full}$). The left panel shows SN-only constraints, while
the right panel shows constraints when BAO and CMB information is also
included.

The figure of merit (FoM) for this model defined by the Dark Energy Task Force
(DETF) \cite{DETF,Huterer_Turner} is the inverse of the area of the 95.4\%
confidence level region $A_{95}$ in the $w_0$--$w_a$ plane. To be precise
about its normalization, we simply define the FoM as in \citet{MHH_FoM} as
\begin{equation}
\mathrm{FoM}^{(w_0 \, w_a)} \equiv (\det \mathbf{C})^{-1/2} \approx \frac{6.17 \pi}{A_{95}}.
\label{eq:fomw0wa}
\end{equation}
The approximate equality in Eq.~\eqref{eq:fomw0wa} becomes exact for a
Gaussian posterior distribution. The FoMs for various scenarios in the
$w_0$--$w_a$ plane are given in Table~\ref{tab:w0wafom}. We find that
including the systematic errors reduces the FoM by about a factor of two to
three.

\begin{figure}[t]
\includegraphics[width=0.35\textwidth]{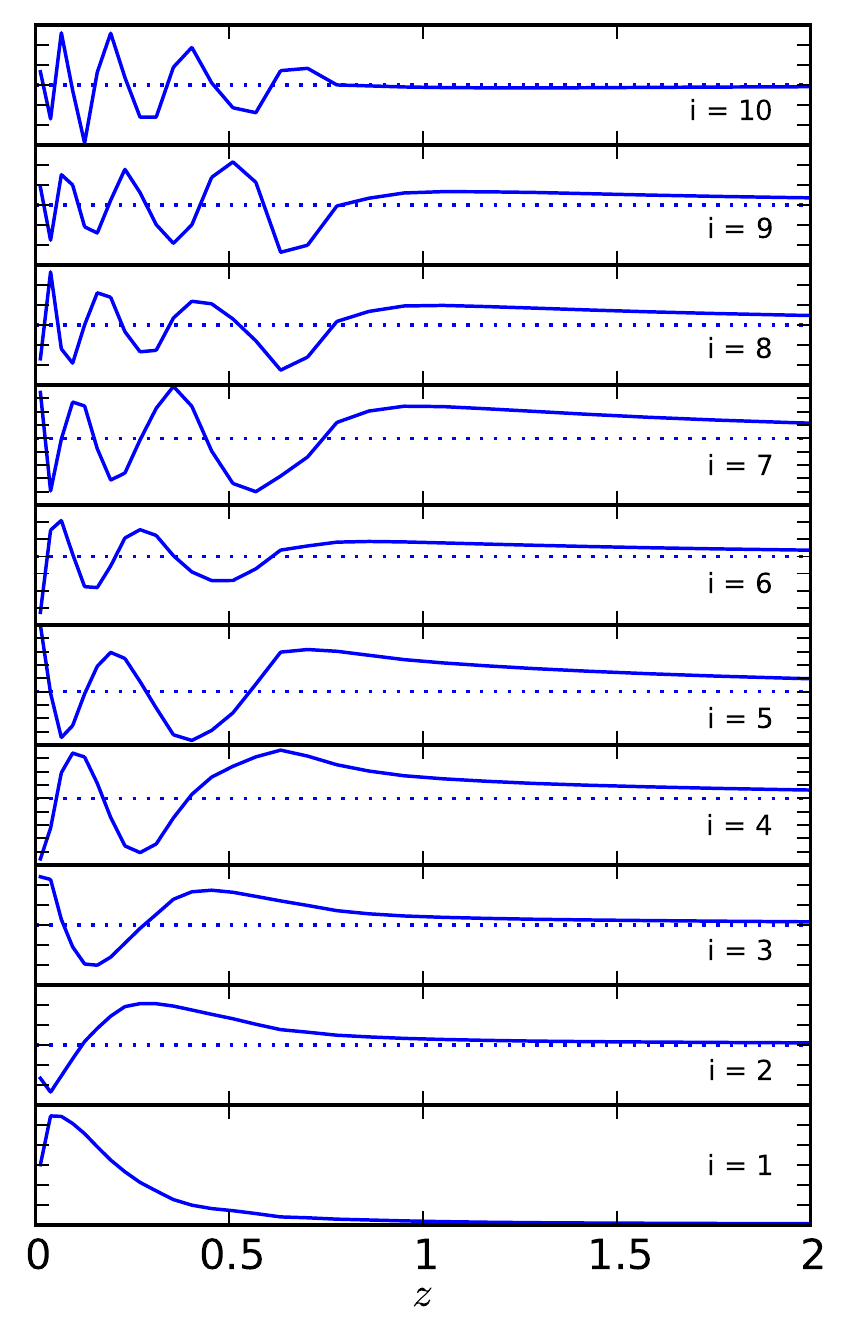}
\caption{The first 10 PCs, $e_1(z)$--$e_{10}(z)$, used in our
  analysis, in order of increasing variance (bottom to top). The PCs
  were obtained assuming the observable quantities centered at the
  fiducial $\Lambda \mathrm{CDM}$ model, but with actual errors from
  the current data. See text for details.}
\label{fig:PCs}
\end{figure}

\begin{table}[h]
\begin{center}
\setlength{\tabcolsep}{0.7em}
\begin{tabular}{|| c | c | c ||}
\hline\hline
\rule[-3mm]{0mm}{8mm} $\mathrm{FoM}^{(w_0 \, w_a)}$ & $\mathbf{D}^\mathrm{stat}$ & $\mathbf{C}^\mathrm{full}$ \\ \hline\hline
\rule[-3mm]{0mm}{8mm} SN & 2.28 & 1.16 \\ \hline
\rule[-3mm]{0mm}{8mm} SN+BAO+CMB & 32.9 & 11.8 \\ \hline
\end{tabular}
\caption{Values of the DETF FoM for SN alone (middle row) and SN+BAO+CMB
  (bottom row). The middle column shows the FoMs for the diagonal covariance
  matrix $\mathbf{D}^\mathrm{stat}$ only, while the right column shows the
  FoMs for the full covariance matrix $\mathbf{C}^\mathrm{full}$. Note that
  including the systematics reduces the FoM by a factor of two to three.}
\label{tab:w0wafom}
\end{center}
\end{table}

\subsection{Principal Components}

We now describe the methodology of how to calculate and constrain the
principal components of DE \cite{Huterer_Starkman}, which are weights
in redshift ordered by how well they are measured by a given
cosmological probe and with a given survey.

Following e.g.\ \citet{Mort_current}, we first precompute the PCs assuming the
current data {\it centered at a fixed fiducial model} (we choose the standard
flat $\Lambda \mathrm{CDM}$ model with $\Omega_M = 1-\Omega_\Lambda =
0.25$). We follow the procedure set forth by the Figure of Merit Science
Working Group (FoMSWG) \cite{FoMSWG} and parametrize $w(z)$ by 36 piecewise
constant values in bins uniformly spaced in scale factor $a$ in the range $0.1
\leq a \leq 1.0$. We fix (i.e.\ ignore) all other parameters in the FoMSWG
except for $\Omega_M$ and the SN Ia nuisance parameter\footnote{In the Fisher
  matrix precomputation of the PCs we assume a single $\mathcal{M}$ parameter
  as per usual practice (and following the FoMSWG parametrization), but in the
  actual constraints on the cosmological parameters we adopt {\it two} such
  parameters as described in Sec.~\ref{sec:data_SN}. To the extent that the
  PCs will be correlated anyway due to the differences between real data and
  assumed ``data'' going into the Fisher matrix, this subtle difference will
  be unimportant.} $\mathcal{M}$ because they are not probed by the SN Ia
data, and at the same time they are effectively marginalized over in the BAO
and CMB data in the distilled observable quantities, $A(z)$ and $R$
respectively, that we use. We fix curvature to zero.

We therefore have a $38 \times 38$ Fisher matrix (or really a $45 \times 45$
Fisher matrix with seven parameters fixed), corresponding to parameters
\begin{equation}
p_i \in \{w_1, \ldots, w_{36}, \Omega_M, \mathcal{M} \}.
\end{equation}
We marginalize over $\Omega_M$ and $\mathcal{M}$ and then diagonalize the remaining $36$-dimensional Fisher
matrix of the piecewise constant $w$ parameters. The resulting eigenvectors --
shapes that describe $w(z)$ -- are the PCs $e_i(z)$, and we
show the 10 best-determined of these PCs, $e_1(z)$--$e_{10}(z)$, in Fig.~\ref{fig:PCs}.

The equation of state can be described as \cite{Mort_falsifying}
\begin{equation}
1 + w(z) = \sum_{i = 1}^N \alpha_i e_i(z),
\end{equation}
where $\alpha_i$ are amplitudes for each PC $e_i(z)$. While the Fisher matrix
tells us the best accuracy to which these PCs are measured using the assumed
data set (these accuracies are related to the eigenvalues $\lambda_i$ via
$\sigma(\alpha_i) = \lambda_i^{-1/2}$), we are not interested in this; rather,
we would like to constrain the PCs using actual current data.

We then feed the shapes in redshift of the first several PCs to the MCMC
procedure to constrain these (and a few other, non-$w(z)$) parameters.

Finally, in our parameter search we impose weak priors on the PCs. Following
\cite{Mort_falsifying} we impose a hard-bound prior on each $\alpha_i$,
enforcing its contribution to excursions in the equation of state to the
region $|1 + w(z)| \leq 1$. This approach yields top-hat priors of
width \cite{MHH_FoM}
\begin{equation}
\Delta \alpha_i = \frac{2}{N_{z,\mathrm{PC}}} \sum_{j=1}^{N_{z,\mathrm{PC}}} | e_i(z_j)|
\label{eq:deltaalpha}
\end{equation}
centered at $w(z) = -1$ or $\alpha_i = 0$. As we will demonstrate, these
priors are much wider than the allowed ranges for most of the individual PCs,
meaning that our principal results are largely unaffected by the prior (Indeed, we verified this explicitly by constraining the PCs without the prior).

\begin{figure*}
\includegraphics[width=\textwidth]{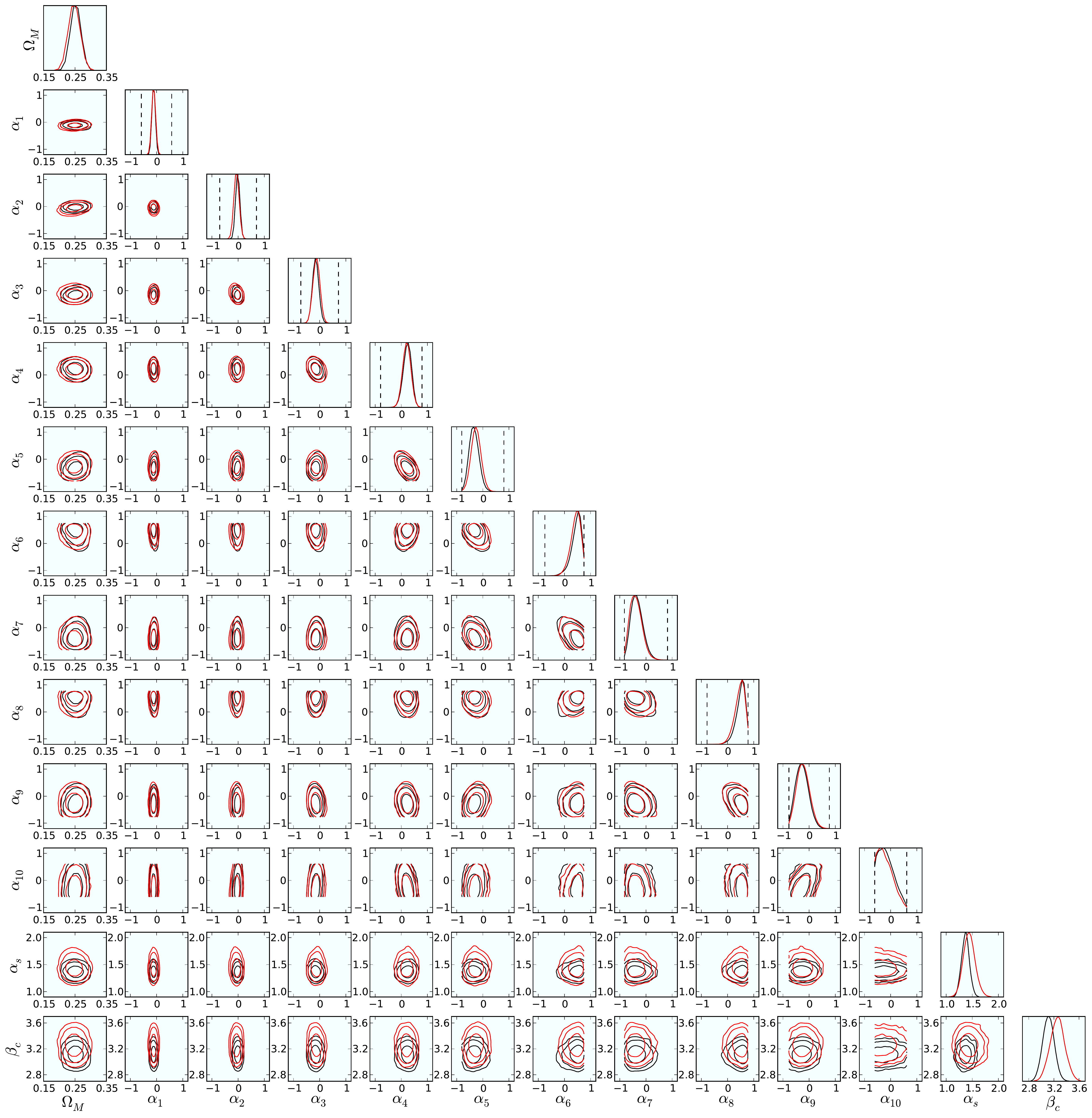}
\caption{68.3\%, 95.4\%, and 99.7\% likelihood contours for all
  pairwise combinations of the 13 cosmological parameters. Diagonal
  boxes show the 1-D marginalized likelihood for each parameter. The
  black contours illustrate the case of diagonal statistical SN errors
  only ($\mathbf{D}^\mathrm{stat}$), while the red contours include
  the full systematic SN covariance matrix
  ($\mathbf{C}^\mathrm{full}$). The parameter ordering is (top to
  bottom, or left to right): matter density relative to critical
  $\Omega_M$, the 10 PC amplitudes $\alpha_1$--$\alpha_{10}$, and the
  stretch and color nuisance parameters $\alpha_s$ and $\beta_c$. Note
  the good constraints on all parameters except for the last few PC amplitudes.}
\label{fig:gridPCs}
\end{figure*}

\begin{figure*}[t]
\includegraphics[width=\textwidth]{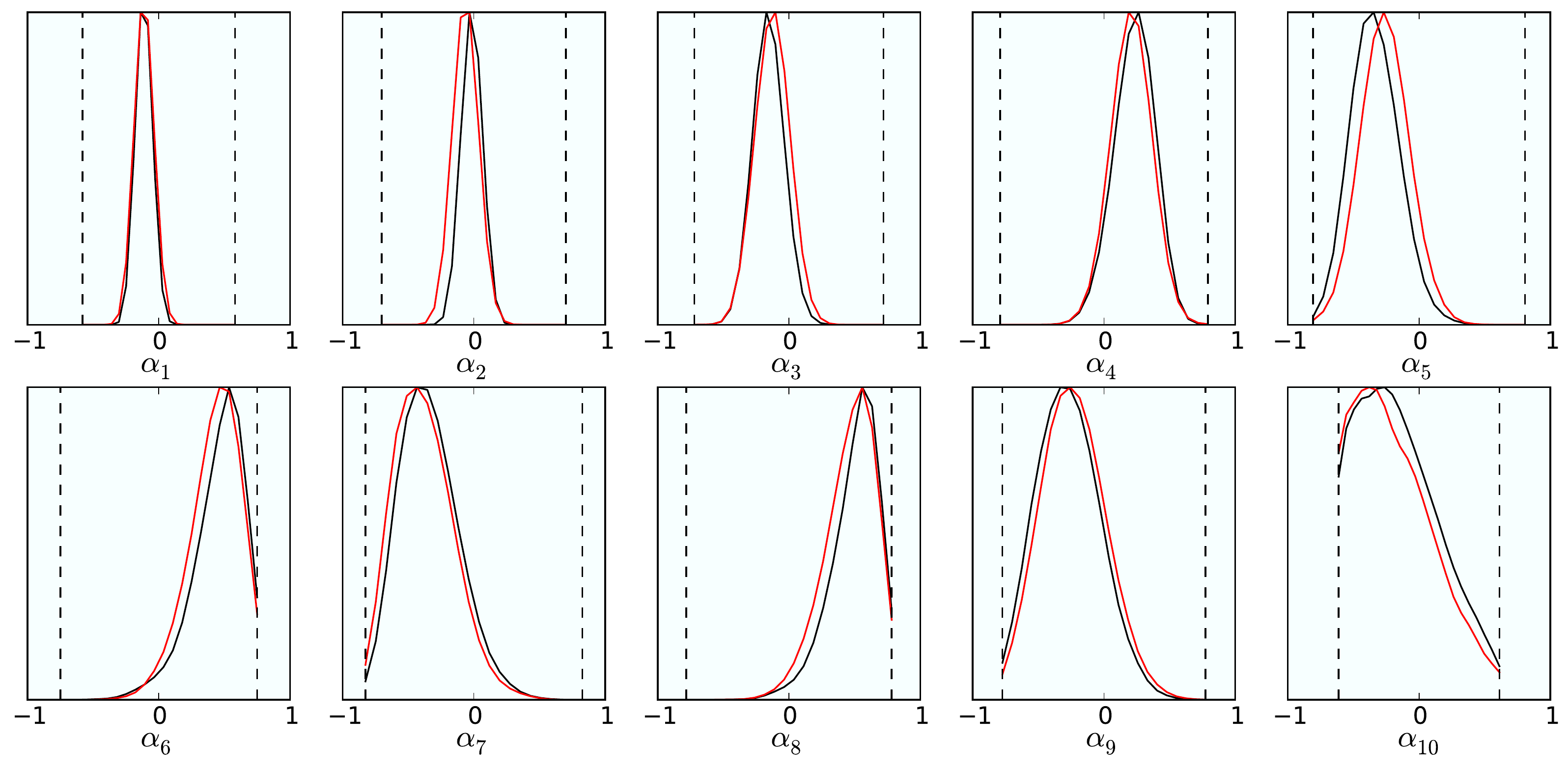}
\caption{Marginalized SN+BAO+CMB constraints on the 10 PC amplitudes. The
  dashed vertical lines represent the prior limits. Black curves represent
  constraints from the diagonal statistical SN errors only, while red curves
  correspond to the full SN covariance matrix. Note the good constraints on
  all PC amplitudes except for the last few.}
\label{fig:PCconstraints}
\end{figure*}

The pairwise constraints on all 13 parameters ($\Omega_M$, the PC amplitudes
$\alpha_1 - \alpha_{10}$, and the nuisance parameters $\alpha_s$ and
$\beta_c$) are shown in Fig.~\ref{fig:gridPCs}. The black curves represent
constraints from the diagonal statistical SN errors only, while the red curves
correspond to the full SN covariance matrix. Overall, the systematic errors
broaden and shift the contours slightly.

In Fig.~\ref{fig:PCconstraints}, we show the individual marginalized
constraints on the 10 PC amplitudes. We are extremely encouraged by the fact
that constraints on more than five of the lowest PCs are very good even with
current data, a result incidentally also found by \cite{Mort_current} using a
slightly different combined ``current'' data set that, most notably, did not
include the BOSS and WiggleZ BAO measurements. Here we again see that the SN systematics broaden the constraints slightly; however,
as we show just below, the cumulative effect of the systematics on the FoM is
not negligible.

\begin{figure}[t]
\includegraphics[width=0.45\textwidth]{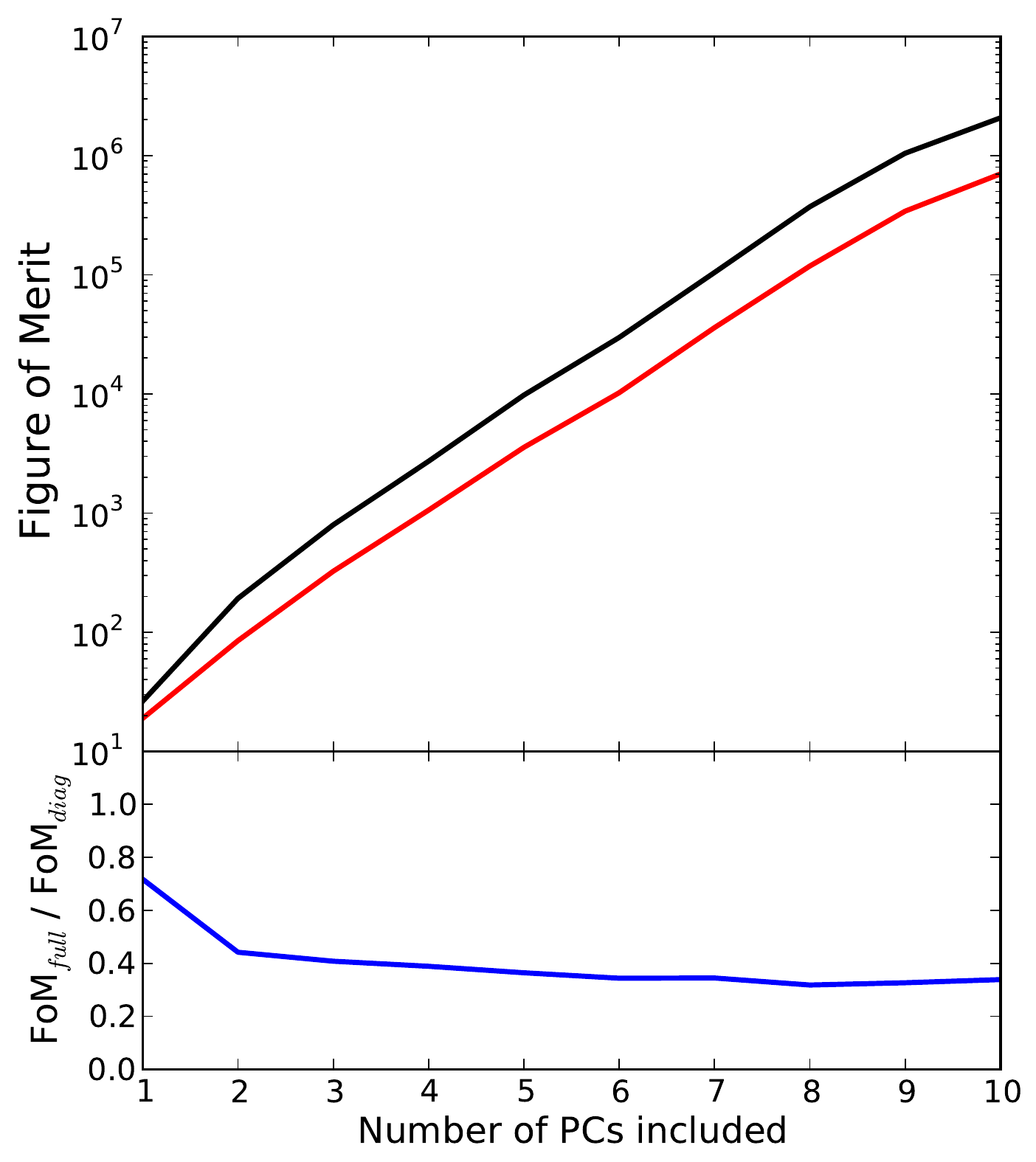}
\caption{Top panel: FoM as a function of the number of PCs included,
  with the black line showing the statistical-only FoM and the red
  line showing the FoM with systematics included (See
  Eq.~\eqref{eq:FoM} for the definition of the FoM). Bottom panel:
  ratio of the FoM with systematic errors considered in the SN Ia data
  to that with only statistical errors considered. BAO and CMB
  constraints were included in both cases. Notice that the FoM ratio
  levels off after approximately two PCs have been added to the
  analysis.}
\label{fig:fomvary}
\end{figure}

We finally calculate the generalization of the DETF FoM to PCs. As defined in \citet{MHH_FoM},
\begin{equation}
\mathrm{FoM}_n^{(\mathrm{PC})} \equiv \left(\frac{\det \mathbf{C}_n}{\det \mathbf{C}_n^{(\mathrm{prior})}}\right)^{-1/2},
\label{eq:FoM}
\end{equation}
where $\mathbf{C}_n$ is the $n \times n$ covariance submatrix of $n$ PCs and 
\begin{equation*}
\det \mathbf{C}_n^{(\mathrm{prior})} = \prod_{i = 1}^n\left(\frac{\Delta \alpha_i}{\sqrt{12}}\right)^2
\end{equation*}
is the determinant of the top-hat prior covariance on the $n$ PC
coefficients. Each $(\Delta \alpha_i/ \sqrt{12})^2$ term refers to the RMS value of the top-hat prior, where $\Delta \alpha_i$ is the width of the top-hat
prior as calculated in Eq.~\eqref{eq:deltaalpha}.

FoM results are shown in Fig.~\ref{fig:fomvary}, where we show the FoM as a
function of the number of PCs included. The top panel shows the
FoMs with and without SN systematic errors, while the bottom panel shows the
corresponding ratios of the two cases. We see that the FoM degradation with the
addition of SN systematic errors asymptotes to about a factor of two when only
two PCs are included and after that remains relatively constant. We therefore conclude
that only the lowest two PCs are affected by current systematic errors. We
suspect that this is due to the fact that the effect of the systematics is
smooth in redshift, and therefore systematics do not become degenerate with
the higher PCs that wiggle in $z$ (see the PC shapes in Fig.~\ref{fig:PCs}). It is somewhat fortuitous that higher ($n > 2$) PCs seem to be unaffected by
systematics, since it is precisely those higher PCs that are difficult to
measure accurately; however, it may be the case that systematics in future
data will behave differently and affect the higher components.

\section{Effect of Finite Detection Significance of BAO}\label{sec:BAO_detect}

In an interesting paper, \citet{Bassett_Afshordi} pointed out that for
marginal detections of cosmological observable quantities, a Gaussian assumption
for the likelihood may be a poor one, especially for models that are several--$\sigma$ away from the central value of the observed quantity. This happens
because the usual Gaussian likelihood implicitly ignores the possibility that
the observed quantity has {\it not} actually been detected in the data at
all. That possibility may have non-negligible probability, and in that
case a flat likelihood in the observable may be more appropriate. In other
words, writing a total likelihood of parameters $\mathbf{p}$ as a function of
data vector $\mathbf{d}$, we have
\begin{align}
P(\mathbf{p}|\mathbf{d}) &= P_\mathrm{detect} \, P(\mathbf{p}|\mathbf{d},\mathrm{detect}) \label{eq:P_detect_noise} \\[0.2cm]
&+ (1 - P_\mathrm{detect}) \, P(\mathbf{p}|\mathbf{d},\mathrm{noise}) \nonumber
\end{align}
where $P_\mathrm{detect}$ is the probability that the obsevable quantity
has actually been detected and $P(\mathbf{p}|\mathbf{d},\mathrm{detect})$
is the likelihood of the cosmological parameters in that case. The
cosmological parameter likelihood $P(\mathbf{p}|\mathbf{d},\mathrm{noise})$ corresponds to the case that the
observable feature was actually noise, and it can be represented by a flat
distribution in the parameters $\mathbf{p}$. {\it Most BAO analyses effectively assume that $P_\mathrm{detect} = 1$}, thus ignoring the
higher-than-expected tail in the overall likelihood coming from the nonzero
second term on the right-hand side of Eq.~\eqref{eq:P_detect_noise}. If the
BAO feature has been detected at very high significance then this is a good
assumption, but it is not {\it a priori} clear that this is the case with all
of the current BAO surveys which typically have several--$\sigma$ detection
significances.

To account for the diminished power of the observations to discriminate
between cosmological models when detection significance is not high,
\citet{Bassett_Afshordi} suggest a fitting function which replaces the usual
Gaussian $\chi^2$ expression $\Delta \chi_G^2$ with
\begin{equation}
\Delta \chi^2 = \frac{\Delta\chi_G^2}
{\sqrt{1 + \left(\displaystyle\frac{\mathrm{S}}{\mathrm{N}}\right)^{-4} \,\Delta\chi_G^4}},
\end{equation}
where $\mathrm{S}/\mathrm{N}$ is the signal-to-noise ratio or detection
significance of the observable feature or quantity. With this
prescription, the quantity $\Delta \chi^2$ is equal to its Gaussian
counterpart for departures from the best-fit model that are small compared to
the signal-to-noise of the observed feature, but it asymptotes to a constant ``tail'' $(\mathrm{S}/\mathrm{N})^2$ in the opposite limit, when $\Delta \chi_G^2 \gg (\mathrm{S}/\mathrm{N})^2$.

Here we apply this reasoning to the measurement of the BAO feature. The
significances of the detection of the BAO feature are 2.4$\sigma$
(corresponding to $\mathrm{S}/\mathrm{N} = 2.4$) for 6dF \cite{Beutler:2011hx},
2.8$\sigma$ for WiggleZ \cite{Blake:2011en} (combined for three redshift
bins), 3.6$\sigma$ for SDSS \cite{Percival:2009xn} (combined for two redshift
bins), and 5.0$\sigma$ for BOSS \cite{Anderson:2012sa}. We expect that, once
the probability of non-detection of the BAO feature has been included, the BAO
constraints will change, especially for surveys with lower significance of
detection and for 99.7\% contour regions. This has in fact been confirmed by
\citet{Bassett_Afshordi} for the case of the SDSS BAO data alone.

\begin{figure*}[t]
\includegraphics[width=\textwidth]{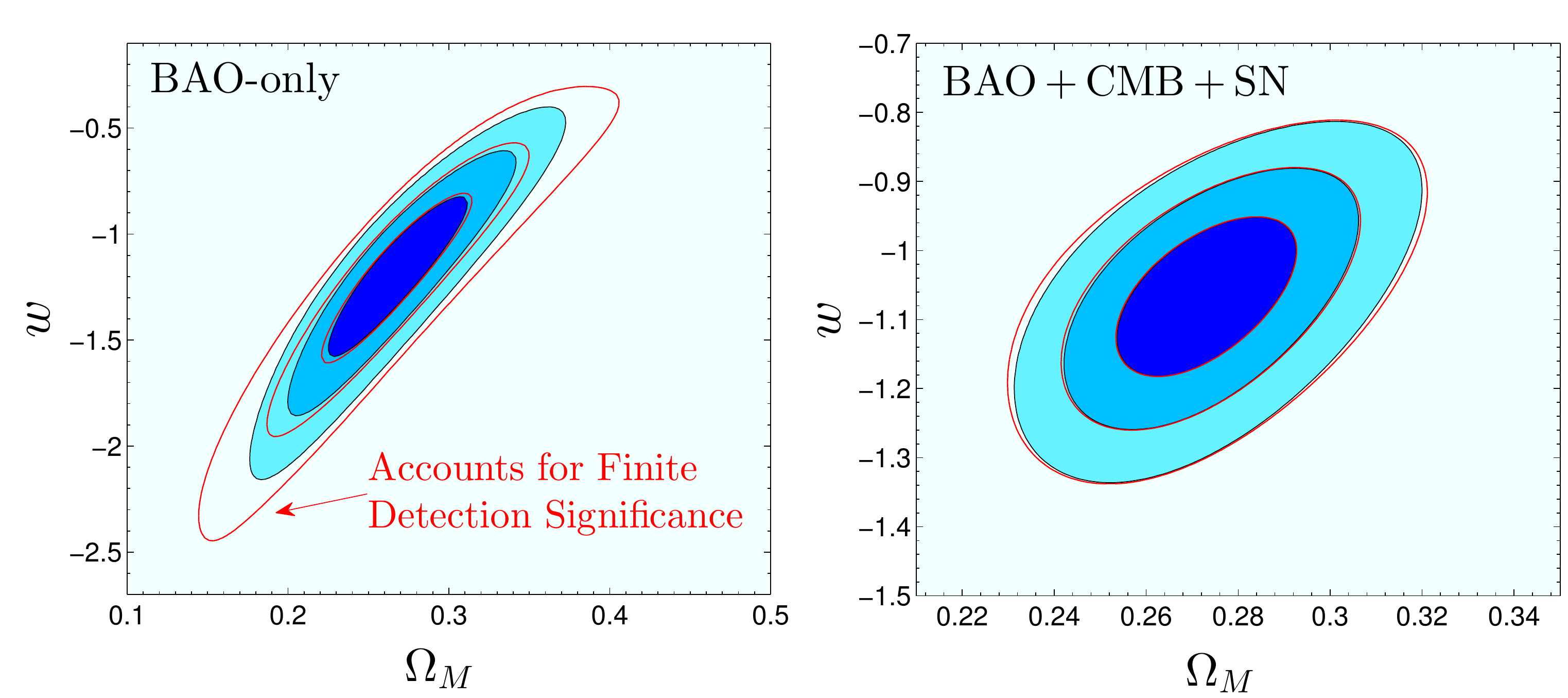}
\caption{Effects on the BAO-only (left panel) and BAO+CMB+SN (right panel)
  constraints in the $\Omega_M$--$w$ plane with (red) and without (shaded blue) the finite
  detection significances of the BAO features taken into account. Note that the differences
  are modest in the BAO-only case and negligible in the combined case.}
\label{fig:BAO_nondetect}
\end{figure*}

Fig.~\ref{fig:BAO_nondetect} shows the effects on the BAO-only (left panel)
and BAO+CMB+SN (right panel) constraints in the $\Omega_M$--$w$ plane with and
without the finite detection of the BAO features taken into
account\footnote{The results in the $w_0$--$w_a$ plane are qualitatively
  similar, and we do not show them here.}. Note that the differences are
modest in the BAO-only case and negligible in the combined case. This is as
expected, especially given that some of the strongest BAO data sets
(e.g.\ BOSS) also have the highest detection significances of the BAO feature.

Note also that there is nothing BAO-specific to the effects of the finite
detection significance. While the CMB is detected with very high confidence
and thus does not warrant a similar analysis, it could be applied to SNe Ia
where, for example, a few percent of SNe may not be Type
Ia\footnote{\citet{Conley2011} find that the fraction of
  non-Ia SNe rises from zero at low redshift to $O(10\%)$ at $z\sim
  1$; however, their modeling is very conservative, and the true fraction of
  non-Ia SNe is likely very small in the current data sets.}. Given the full
probabilistic classification of each SN on whether or not it is Type Ia
\cite{Kessler_photo_classification,BEAMS}, one could carry out a similar
analysis, which in this context would be how imperfect purity of the SN Ia
sample affects the constraints on cosmological parameters. We suspect the
results would be even less discrepant relative to the usual perfect-detection
analysis than in the case of BAO, and we do not pursue such an analysis in
this paper.

In conclusion, the finite detection significance of the BAO feature in
large-scale structure surveys leads to a small but discernible weakening of
the constraints on cosmological parameters.

\section{Conclusions}\label{sec:conclude}

In this paper, we have investigated the effects of systematic errors in
current SN Ia observations on DE parameter constraints. We accounted for the
systematic errors in SN Ia observations, including the effects of photometric
calibration, dust, color, gravitational lensing, and other systematics by
adopting a fully off-diagonal covariance matrix between $\sim 500$ SNe from
the SNLS compilation (see Fig.~\ref{fig:covmatrixplot}). We extended the
similar analysis from \citet{Conley2011} by constraining the temporal
evolution of the equation of state of DE described by the pair of parameters
$(w_0, w_a)$ as well as a much richer description in terms of 10 PCs of the
equation of state (shown in Fig.~\ref{fig:PCs}). We combined the SN Ia
constraints with data from BAO from four different surveys (see
Fig.~\ref{fig:Ameas}) as well as the principal information on DE given by the
acoustic peak measurements of the CMB anisotropies measured by the WMAP experiment.

The constraints on the simple parametrizations of DE are affected by the
systematics, but the overall constraints are still strong even after their
inclusion (see Figs.~\ref{fig:Omw} and \ref{fig:w0wa}). More importantly, we
found that systematic errors affect the contraints somewhat, reducing the DETF
FoM by a factor of about three (see Table~\ref{tab:w0wafom}), while the
generalized PC-based FoM is degraded by a factor of two (see
Fig.~\ref{fig:fomvary}). However, as the PC analysis shows, this degradation
is mainly restricted to the first two numbers (PC amplitudes) describing
DE. In fact, what is particularly impressive about current data is that more than five PCs are well-constrained even in the presence of systematic errors (see Figs.~\ref{fig:gridPCs} and \ref{fig:PCconstraints}).

In the spirit of testing for systematic effects in current data constraining
DE, we also wondered if the relatively low detection significances of BAO
features, ranging from about 2.4-$\sigma$ to 5.0-$\sigma$ in various
surveys, change the overall cosmological constraints. While not a systematic
error {\it per se}, a small but non-negligible probability that the BAO
feature has not been detected in some of these surveys implies that the
posterior probability of cosmological parameter values asymptotes to a small
but nonzero value far from the likelihood peak \cite{Bassett_Afshordi}. We
find that, while the BAO-only constraints are somewhat affected, the combined
constraints are not (see Fig.~\ref{fig:BAO_nondetect}).

From all this, we conclude that current systematic errors do degrade DE
constraints and FoMs, but not in a major way. Given that future constraints
are forecasted to be much better, continued control of current systematic
errors remains key for progress in characterizing DE.

\section{Acknowledgments}
We thank Michael Mortonson and Benedikt Diemer for thoughtful comments on the manuscript.
DH has been supported by the DOE, NASA and the NSF. EJR and DLS
thank the Santa Fe Cosmology Workshop for hospitality, while DH thanks the
Aspen Center for Physics, which is supported by the National Science
Foundation Grant No.\ 1066293.

\bibliography{pcfom}

\begin{thebibliography}{51}%
\makeatletter
\providecommand \@ifxundefined [1]{%
 \@ifx{#1\undefined}
}%
\providecommand \@ifnum [1]{%
 \ifnum #1\expandafter \@firstoftwo
 \else \expandafter \@secondoftwo
 \fi
}%
\providecommand \@ifx [1]{%
 \ifx #1\expandafter \@firstoftwo
 \else \expandafter \@secondoftwo
 \fi
}%
\providecommand \natexlab [1]{#1}%
\providecommand \enquote  [1]{``#1''}%
\providecommand \bibnamefont  [1]{#1}%
\providecommand \bibfnamefont [1]{#1}%
\providecommand \citenamefont [1]{#1}%
\providecommand \href@noop [0]{\@secondoftwo}%
\providecommand \href [0]{\begingroup \@sanitize@url \@href}%
\providecommand \@href[1]{\@@startlink{#1}\@@href}%
\providecommand \@@href[1]{\endgroup#1\@@endlink}%
\providecommand \@sanitize@url [0]{\catcode `\\12\catcode `\$12\catcode
  `\&12\catcode `\#12\catcode `\^12\catcode `\_12\catcode `\%12\relax}%
\providecommand \@@startlink[1]{}%
\providecommand \@@endlink[0]{}%
\providecommand \url  [0]{\begingroup\@sanitize@url \@url }%
\providecommand \@url [1]{\endgroup\@href {#1}{\urlprefix }}%
\providecommand \urlprefix  [0]{URL }%
\providecommand \Eprint [0]{\href }%
\providecommand \doibase [0]{http://dx.doi.org/}%
\providecommand \selectlanguage [0]{\@gobble}%
\providecommand \bibinfo  [0]{\@secondoftwo}%
\providecommand \bibfield  [0]{\@secondoftwo}%
\providecommand \translation [1]{[#1]}%
\providecommand \BibitemOpen [0]{}%
\providecommand \bibitemStop [0]{}%
\providecommand \bibitemNoStop [0]{.\EOS\space}%
\providecommand \EOS [0]{\spacefactor3000\relax}%
\providecommand \BibitemShut  [1]{\csname bibitem#1\endcsname}%
\let\auto@bib@innerbib\@empty
\bibitem [{\citenamefont {Riess}\ \emph {et~al.}(1998)\citenamefont {Riess}
  \emph {et~al.}}]{Riess_1998}%
  \BibitemOpen
  \bibfield  {author} {\bibinfo {author} {\bibfnamefont {A.~G.}\ \bibnamefont
  {Riess}} \emph {et~al.},\ }\href@noop {} {\bibfield  {journal} {\bibinfo
  {journal} {Astron. J.}\ }\textbf {\bibinfo {volume} {116}},\ \bibinfo {pages}
  {1009} (\bibinfo {year} {1998})},\ \Eprint
  {http://arxiv.org/abs/astro-ph/9805201} {astro-ph/9805201} \BibitemShut
  {NoStop}%
\bibitem [{\citenamefont {Perlmutter}\ \emph {et~al.}(1999)\citenamefont
  {Perlmutter} \emph {et~al.}}]{Perlmutter_1999}%
  \BibitemOpen
  \bibfield  {author} {\bibinfo {author} {\bibfnamefont {S.}~\bibnamefont
  {Perlmutter}} \emph {et~al.},\ }\href@noop {} {\bibfield  {journal} {\bibinfo
   {journal} {Astrophys. J.}\ }\textbf {\bibinfo {volume} {517}},\ \bibinfo
  {pages} {565} (\bibinfo {year} {1999})},\ \Eprint
  {http://arxiv.org/abs/astro-ph/9812133} {astro-ph/9812133} \BibitemShut
  {NoStop}%
\bibitem [{\citenamefont {Frieman}\ \emph {et~al.}(2008)\citenamefont
  {Frieman}, \citenamefont {Turner},\ and\ \citenamefont
  {Huterer}}]{FriTurHut}%
  \BibitemOpen
  \bibfield  {author} {\bibinfo {author} {\bibfnamefont {J.}~\bibnamefont
  {Frieman}}, \bibinfo {author} {\bibfnamefont {M.}~\bibnamefont {Turner}}, \
  and\ \bibinfo {author} {\bibfnamefont {D.}~\bibnamefont {Huterer}},\ }\href
  {\doibase 10.1146/annurev.astro.46.060407.145243} {\bibfield  {journal}
  {\bibinfo  {journal} {Ann.\ Rev.\ Astr.\ Astrophys.}\ }\textbf {\bibinfo
  {volume} {46}},\ \bibinfo {pages} {385} (\bibinfo {year} {2008})},\ \Eprint
  {http://arxiv.org/abs/0803.0982} {arXiv:0803.0982} \BibitemShut {NoStop}%
\bibitem [{\citenamefont {Weinberg}\ \emph {et~al.}(2012)\citenamefont
  {Weinberg}, \citenamefont {Mortonson}, \citenamefont {Eisenstein},
  \citenamefont {Hirata}, \citenamefont {Riess} \emph
  {et~al.}}]{Weinberg:2012es}%
  \BibitemOpen
  \bibfield  {author} {\bibinfo {author} {\bibfnamefont {D.~H.}\ \bibnamefont
  {Weinberg}}, \bibinfo {author} {\bibfnamefont {M.~J.}\ \bibnamefont
  {Mortonson}}, \bibinfo {author} {\bibfnamefont {D.~J.}\ \bibnamefont
  {Eisenstein}}, \bibinfo {author} {\bibfnamefont {C.}~\bibnamefont {Hirata}},
  \bibinfo {author} {\bibfnamefont {A.~G.}\ \bibnamefont {Riess}},  \emph
  {et~al.},\ }\href@noop {} {\  (\bibinfo {year} {2012})},\ \Eprint
  {http://arxiv.org/abs/1201.2434} {arXiv:1201.2434} \BibitemShut {NoStop}%
\bibitem [{\citenamefont {Conley}\ \emph {et~al.}(2011)\citenamefont {Conley},
  \citenamefont {Guy}, \citenamefont {Sullivan}, \citenamefont {Regnault},
  \citenamefont {Astier} \emph {et~al.}}]{Conley2011}%
  \BibitemOpen
  \bibfield  {author} {\bibinfo {author} {\bibfnamefont {A.}~\bibnamefont
  {Conley}}, \bibinfo {author} {\bibfnamefont {J.}~\bibnamefont {Guy}},
  \bibinfo {author} {\bibfnamefont {M.}~\bibnamefont {Sullivan}}, \bibinfo
  {author} {\bibfnamefont {N.}~\bibnamefont {Regnault}}, \bibinfo {author}
  {\bibfnamefont {P.}~\bibnamefont {Astier}},  \emph {et~al.},\ }\href
  {\doibase 10.1088/0067-0049/192/1/1} {\bibfield  {journal} {\bibinfo
  {journal} {Astrophys.J.Suppl.}\ }\textbf {\bibinfo {volume} {192}},\ \bibinfo
  {pages} {1} (\bibinfo {year} {2011})},\ \Eprint
  {http://arxiv.org/abs/1104.1443} {arXiv:1104.1443} \BibitemShut {NoStop}%
\bibitem [{\citenamefont {Wood-Vasey}\ \emph {et~al.}(2007)\citenamefont
  {Wood-Vasey} \emph {et~al.}}]{WoodVasey_2007}%
  \BibitemOpen
  \bibfield  {author} {\bibinfo {author} {\bibfnamefont {W.~M.}\ \bibnamefont
  {Wood-Vasey}} \emph {et~al.},\ }\href {\doibase 10.1086/518642} {\bibfield
  {journal} {\bibinfo  {journal} {Astrophys. J.}\ }\textbf {\bibinfo {volume}
  {666}},\ \bibinfo {pages} {694} (\bibinfo {year} {2007})},\ \Eprint
  {http://arxiv.org/abs/astro-ph/0701041} {astro-ph/0701041} \BibitemShut
  {NoStop}%
\bibitem [{\citenamefont {Hicken}\ \emph {et~al.}(2009)\citenamefont {Hicken}
  \emph {et~al.}}]{Constitution}%
  \BibitemOpen
  \bibfield  {author} {\bibinfo {author} {\bibfnamefont {M.}~\bibnamefont
  {Hicken}} \emph {et~al.},\ }\href {\doibase 10.1088/0004-637X/700/2/1097}
  {\bibfield  {journal} {\bibinfo  {journal} {Astrophys. J.}\ }\textbf
  {\bibinfo {volume} {700}},\ \bibinfo {pages} {1097} (\bibinfo {year}
  {2009})},\ \Eprint {http://arxiv.org/abs/0901.4804} {arXiv:0901.4804}
  \BibitemShut {NoStop}%
\bibitem [{\citenamefont {Kessler}\ \emph {et~al.}(2009)\citenamefont {Kessler}
  \emph {et~al.}}]{SDSS_SN}%
  \BibitemOpen
  \bibfield  {author} {\bibinfo {author} {\bibfnamefont {R.}~\bibnamefont
  {Kessler}} \emph {et~al.},\ }\href {\doibase 10.1088/0067-0049/185/1/32}
  {\bibfield  {journal} {\bibinfo  {journal} {Astrophys. J. Suppl.}\ }\textbf
  {\bibinfo {volume} {185}},\ \bibinfo {pages} {32} (\bibinfo {year} {2009})},\
  \Eprint {http://arxiv.org/abs/0908.4274} {arXiv:0908.4274} \BibitemShut
  {NoStop}%
\bibitem [{\citenamefont {Sullivan}\ \emph {et~al.}(2011)\citenamefont
  {Sullivan}, \citenamefont {Guy}, \citenamefont {Conley}, \citenamefont
  {Regnault}, \citenamefont {Astier} \emph {et~al.}}]{Sullivan2011}%
  \BibitemOpen
  \bibfield  {author} {\bibinfo {author} {\bibfnamefont {M.}~\bibnamefont
  {Sullivan}}, \bibinfo {author} {\bibfnamefont {J.}~\bibnamefont {Guy}},
  \bibinfo {author} {\bibfnamefont {A.}~\bibnamefont {Conley}}, \bibinfo
  {author} {\bibfnamefont {N.}~\bibnamefont {Regnault}}, \bibinfo {author}
  {\bibfnamefont {P.}~\bibnamefont {Astier}},  \emph {et~al.},\ }\href
  {\doibase 10.1088/0004-637X/737/2/102} {\bibfield  {journal} {\bibinfo
  {journal} {Astrophys.J.}\ }\textbf {\bibinfo {volume} {737}},\ \bibinfo
  {pages} {102} (\bibinfo {year} {2011})},\ \Eprint
  {http://arxiv.org/abs/1104.1444} {arXiv:1104.1444} \BibitemShut {NoStop}%
\bibitem [{\citenamefont {Davis}\ \emph {et~al.}(2007)\citenamefont {Davis},
  \citenamefont {Mortsell}, \citenamefont {Sollerman}, \citenamefont {Becker},
  \citenamefont {Blondin} \emph {et~al.}}]{Davetal07}%
  \BibitemOpen
  \bibfield  {author} {\bibinfo {author} {\bibfnamefont {T.~M.}\ \bibnamefont
  {Davis}}, \bibinfo {author} {\bibfnamefont {E.}~\bibnamefont {Mortsell}},
  \bibinfo {author} {\bibfnamefont {J.}~\bibnamefont {Sollerman}}, \bibinfo
  {author} {\bibfnamefont {A.}~\bibnamefont {Becker}}, \bibinfo {author}
  {\bibfnamefont {S.}~\bibnamefont {Blondin}},  \emph {et~al.},\ }\href
  {\doibase 10.1086/519988} {\bibfield  {journal} {\bibinfo  {journal}
  {Astrophys.J.}\ }\textbf {\bibinfo {volume} {666}},\ \bibinfo {pages} {716}
  (\bibinfo {year} {2007})},\ \Eprint {http://arxiv.org/abs/astro-ph/0701510}
  {astro-ph/0701510} \BibitemShut {NoStop}%
\bibitem [{\citenamefont {Rubin}\ \emph {et~al.}(2009)\citenamefont {Rubin},
  \citenamefont {Linder}, \citenamefont {Kowalski}, \citenamefont {Aldering},
  \citenamefont {Amanullah} \emph {et~al.}}]{Rubin_SCP}%
  \BibitemOpen
  \bibfield  {author} {\bibinfo {author} {\bibfnamefont {D.}~\bibnamefont
  {Rubin}}, \bibinfo {author} {\bibfnamefont {E.}~\bibnamefont {Linder}},
  \bibinfo {author} {\bibfnamefont {M.}~\bibnamefont {Kowalski}}, \bibinfo
  {author} {\bibfnamefont {G.}~\bibnamefont {Aldering}}, \bibinfo {author}
  {\bibfnamefont {R.}~\bibnamefont {Amanullah}},  \emph {et~al.},\ }\href
  {\doibase 10.1088/0004-637X/695/1/391} {\bibfield  {journal} {\bibinfo
  {journal} {Astrophys.J.}\ }\textbf {\bibinfo {volume} {695}},\ \bibinfo
  {pages} {391} (\bibinfo {year} {2009})},\ \Eprint
  {http://arxiv.org/abs/0807.1108} {arXiv:0807.1108} \BibitemShut {NoStop}%
\bibitem [{\citenamefont {Amanullah}\ \emph {et~al.}(2010)\citenamefont
  {Amanullah}, \citenamefont {Lidman}, \citenamefont {Rubin}, \citenamefont
  {Aldering}, \citenamefont {Astier} \emph {et~al.}}]{Amanullah_Union2}%
  \BibitemOpen
  \bibfield  {author} {\bibinfo {author} {\bibfnamefont {R.}~\bibnamefont
  {Amanullah}}, \bibinfo {author} {\bibfnamefont {C.}~\bibnamefont {Lidman}},
  \bibinfo {author} {\bibfnamefont {D.}~\bibnamefont {Rubin}}, \bibinfo
  {author} {\bibfnamefont {G.}~\bibnamefont {Aldering}}, \bibinfo {author}
  {\bibfnamefont {P.}~\bibnamefont {Astier}},  \emph {et~al.},\ }\href
  {\doibase 10.1088/0004-637X/716/1/712} {\bibfield  {journal} {\bibinfo
  {journal} {Astrophys.J.}\ }\textbf {\bibinfo {volume} {716}},\ \bibinfo
  {pages} {712} (\bibinfo {year} {2010})},\ \Eprint
  {http://arxiv.org/abs/1004.1711} {arXiv:1004.1711} \BibitemShut {NoStop}%
\bibitem [{\citenamefont {Suzuki}\ \emph {et~al.}(2012)\citenamefont {Suzuki},
  \citenamefont {Rubin}, \citenamefont {Lidman}, \citenamefont {Aldering},
  \citenamefont {Amanullah} \emph {et~al.}}]{Suzuki_SCP}%
  \BibitemOpen
  \bibfield  {author} {\bibinfo {author} {\bibfnamefont {N.}~\bibnamefont
  {Suzuki}}, \bibinfo {author} {\bibfnamefont {D.}~\bibnamefont {Rubin}},
  \bibinfo {author} {\bibfnamefont {C.}~\bibnamefont {Lidman}}, \bibinfo
  {author} {\bibfnamefont {G.}~\bibnamefont {Aldering}}, \bibinfo {author}
  {\bibfnamefont {R.}~\bibnamefont {Amanullah}},  \emph {et~al.},\ }\href
  {\doibase 10.1088/0004-637X/746/1/85} {\bibfield  {journal} {\bibinfo
  {journal} {Astrophys.J.}\ }\textbf {\bibinfo {volume} {746}},\ \bibinfo
  {pages} {85} (\bibinfo {year} {2012})},\ \Eprint
  {http://arxiv.org/abs/1105.3470} {arXiv:1105.3470} \BibitemShut {NoStop}%
\bibitem [{\citenamefont {Mortonson}\ \emph
  {et~al.}(2010{\natexlab{a}})\citenamefont {Mortonson}, \citenamefont {Hu},\
  and\ \citenamefont {Huterer}}]{Mort_current}%
  \BibitemOpen
  \bibfield  {author} {\bibinfo {author} {\bibfnamefont {M.~J.}\ \bibnamefont
  {Mortonson}}, \bibinfo {author} {\bibfnamefont {W.}~\bibnamefont {Hu}}, \
  and\ \bibinfo {author} {\bibfnamefont {D.}~\bibnamefont {Huterer}},\
  }\href@noop {} {\bibfield  {journal} {\bibinfo  {journal} {Phys. Rev.}\
  }\textbf {\bibinfo {volume} {D81}},\ \bibinfo {pages} {063007} (\bibinfo
  {year} {2010}{\natexlab{a}})},\ \Eprint {http://arxiv.org/abs/0912.3816}
  {arXiv:0912.3816} \BibitemShut {NoStop}%
\bibitem [{\citenamefont {Mortonson}\ \emph
  {et~al.}(2010{\natexlab{b}})\citenamefont {Mortonson}, \citenamefont {Hu},\
  and\ \citenamefont {Huterer}}]{MHH_FoM}%
  \BibitemOpen
  \bibfield  {author} {\bibinfo {author} {\bibfnamefont {M.}~\bibnamefont
  {Mortonson}}, \bibinfo {author} {\bibfnamefont {W.}~\bibnamefont {Hu}}, \
  and\ \bibinfo {author} {\bibfnamefont {D.}~\bibnamefont {Huterer}},\
  }\href@noop {} {\bibfield  {journal} {\bibinfo  {journal} {Phys. Rev.}\
  }\textbf {\bibinfo {volume} {D82}},\ \bibinfo {pages} {063004} (\bibinfo
  {year} {2010}{\natexlab{b}})}\BibitemShut {NoStop}%
\bibitem [{\citenamefont {Huterer}\ and\ \citenamefont
  {Cooray}(2005)}]{Huterer_Cooray}%
  \BibitemOpen
  \bibfield  {author} {\bibinfo {author} {\bibfnamefont {D.}~\bibnamefont
  {Huterer}}\ and\ \bibinfo {author} {\bibfnamefont {A.}~\bibnamefont
  {Cooray}},\ }\href@noop {} {\bibfield  {journal} {\bibinfo  {journal} {Phys.
  Rev.}\ }\textbf {\bibinfo {volume} {D71}},\ \bibinfo {pages} {023506}
  (\bibinfo {year} {2005})},\ \Eprint {http://arxiv.org/abs/astro-ph/0404062}
  {astro-ph/0404062} \BibitemShut {NoStop}%
\bibitem [{\citenamefont {Wang}\ and\ \citenamefont
  {Tegmark}(2005)}]{Wang_Tegmark_2005}%
  \BibitemOpen
  \bibfield  {author} {\bibinfo {author} {\bibfnamefont {Y.}~\bibnamefont
  {Wang}}\ and\ \bibinfo {author} {\bibfnamefont {M.}~\bibnamefont {Tegmark}},\
  }\href@noop {} {\bibfield  {journal} {\bibinfo  {journal} {Phys. Rev.}\
  }\textbf {\bibinfo {volume} {D71}},\ \bibinfo {pages} {103513} (\bibinfo
  {year} {2005})},\ \Eprint {http://arxiv.org/abs/astro-ph/0501351}
  {astro-ph/0501351} \BibitemShut {NoStop}%
\bibitem [{\citenamefont {Zunckel}\ and\ \citenamefont
  {Trotta}(2007)}]{Zunckel_Trotta}%
  \BibitemOpen
  \bibfield  {author} {\bibinfo {author} {\bibfnamefont {C.}~\bibnamefont
  {Zunckel}}\ and\ \bibinfo {author} {\bibfnamefont {R.}~\bibnamefont
  {Trotta}},\ }\href {\doibase 10.1111/j.1365-2966.2007.12000.x} {\bibfield
  {journal} {\bibinfo  {journal} {Mon. Not. Roy. Astron. Soc.}\ }\textbf
  {\bibinfo {volume} {380}},\ \bibinfo {pages} {865} (\bibinfo {year}
  {2007})},\ \Eprint {http://arxiv.org/abs/astro-ph/0702695} {astro-ph/0702695}
  \BibitemShut {NoStop}%
\bibitem [{\citenamefont {Zhao}\ \emph {et~al.}(2008)\citenamefont {Zhao},
  \citenamefont {Huterer},\ and\ \citenamefont {Zhang}}]{Zhao_Huterer_Zhang}%
  \BibitemOpen
  \bibfield  {author} {\bibinfo {author} {\bibfnamefont {G.-B.}\ \bibnamefont
  {Zhao}}, \bibinfo {author} {\bibfnamefont {D.}~\bibnamefont {Huterer}}, \
  and\ \bibinfo {author} {\bibfnamefont {X.}~\bibnamefont {Zhang}},\ }\href
  {\doibase 10.1103/PhysRevD.77.121302} {\bibfield  {journal} {\bibinfo
  {journal} {Phys. Rev.}\ }\textbf {\bibinfo {volume} {D77}},\ \bibinfo {pages}
  {121302} (\bibinfo {year} {2008})},\ \Eprint {http://arxiv.org/abs/0712.2277}
  {arXiv:0712.2277} \BibitemShut {NoStop}%
\bibitem [{\citenamefont {Hojjati}\ \emph {et~al.}(2010)\citenamefont
  {Hojjati}, \citenamefont {Pogosian},\ and\ \citenamefont
  {Zhao}}]{Hojjati:2009ab}%
  \BibitemOpen
  \bibfield  {author} {\bibinfo {author} {\bibfnamefont {A.}~\bibnamefont
  {Hojjati}}, \bibinfo {author} {\bibfnamefont {L.}~\bibnamefont {Pogosian}}, \
  and\ \bibinfo {author} {\bibfnamefont {G.-B.}\ \bibnamefont {Zhao}},\ }\href
  {\doibase 10.1088/1475-7516/2010/04/007} {\bibfield  {journal} {\bibinfo
  {journal} {JCAP}\ }\textbf {\bibinfo {volume} {1004}},\ \bibinfo {pages}
  {007} (\bibinfo {year} {2010})},\ \Eprint {http://arxiv.org/abs/0912.4843}
  {arXiv:0912.4843} \BibitemShut {NoStop}%
\bibitem [{\citenamefont {Ishida}\ and\ \citenamefont
  {de~Souza}(2011)}]{Ishida:2010nk}%
  \BibitemOpen
  \bibfield  {author} {\bibinfo {author} {\bibfnamefont {E.~E.}\ \bibnamefont
  {Ishida}}\ and\ \bibinfo {author} {\bibfnamefont {R.~S.}\ \bibnamefont
  {de~Souza}},\ }\href {\doibase 10.1051/0004-6361/201015281} {\bibfield
  {journal} {\bibinfo  {journal} {Astron.Astrophys.}\ }\textbf {\bibinfo
  {volume} {527}},\ \bibinfo {pages} {A49} (\bibinfo {year} {2011})},\ \Eprint
  {http://arxiv.org/abs/1012.5335} {arXiv:1012.5335} \BibitemShut {NoStop}%
\bibitem [{\citenamefont {Shafieloo}\ \emph {et~al.}(2012)\citenamefont
  {Shafieloo}, \citenamefont {Kim},\ and\ \citenamefont
  {Linder}}]{Shafieloo:2012ht}%
  \BibitemOpen
  \bibfield  {author} {\bibinfo {author} {\bibfnamefont {A.}~\bibnamefont
  {Shafieloo}}, \bibinfo {author} {\bibfnamefont {A.~G.}\ \bibnamefont {Kim}},
  \ and\ \bibinfo {author} {\bibfnamefont {E.~V.}\ \bibnamefont {Linder}},\
  }\href@noop {} {\  (\bibinfo {year} {2012})},\ \Eprint
  {http://arxiv.org/abs/1204.2272} {arXiv:1204.2272} \BibitemShut {NoStop}%
\bibitem [{\citenamefont {Seikel}\ \emph {et~al.}(2012)\citenamefont {Seikel},
  \citenamefont {Clarkson},\ and\ \citenamefont {Smith}}]{Seikel:2012uu}%
  \BibitemOpen
  \bibfield  {author} {\bibinfo {author} {\bibfnamefont {M.}~\bibnamefont
  {Seikel}}, \bibinfo {author} {\bibfnamefont {C.}~\bibnamefont {Clarkson}}, \
  and\ \bibinfo {author} {\bibfnamefont {M.}~\bibnamefont {Smith}},\ }\href
  {\doibase 10.1088/1475-7516/2012/06/036} {\bibfield  {journal} {\bibinfo
  {journal} {JCAP}\ }\textbf {\bibinfo {volume} {1206}},\ \bibinfo {pages}
  {036} (\bibinfo {year} {2012})},\ \Eprint {http://arxiv.org/abs/1204.2832}
  {arXiv:1204.2832} \BibitemShut {NoStop}%
\bibitem [{\citenamefont {Zhao}\ \emph {et~al.}(2012)\citenamefont {Zhao},
  \citenamefont {Crittenden}, \citenamefont {Pogosian},\ and\ \citenamefont
  {Zhang}}]{Zhao:2012aw}%
  \BibitemOpen
  \bibfield  {author} {\bibinfo {author} {\bibfnamefont {G.-B.}\ \bibnamefont
  {Zhao}}, \bibinfo {author} {\bibfnamefont {R.~G.}\ \bibnamefont
  {Crittenden}}, \bibinfo {author} {\bibfnamefont {L.}~\bibnamefont
  {Pogosian}}, \ and\ \bibinfo {author} {\bibfnamefont {X.}~\bibnamefont
  {Zhang}},\ }\href@noop {} {\  (\bibinfo {year} {2012})},\ \Eprint
  {http://arxiv.org/abs/1207.3804} {arXiv:1207.3804} \BibitemShut {NoStop}%
\bibitem [{\citenamefont {Guy}\ \emph {et~al.}(2007)\citenamefont {Guy},
  \citenamefont {Astier}, \citenamefont {Baumont}, \citenamefont {Hardin},
  \citenamefont {Pain} \emph {et~al.}}]{Guy:2007dv}%
  \BibitemOpen
  \bibfield  {author} {\bibinfo {author} {\bibfnamefont {J.}~\bibnamefont
  {Guy}}, \bibinfo {author} {\bibfnamefont {P.}~\bibnamefont {Astier}},
  \bibinfo {author} {\bibfnamefont {S.}~\bibnamefont {Baumont}}, \bibinfo
  {author} {\bibfnamefont {D.}~\bibnamefont {Hardin}}, \bibinfo {author}
  {\bibfnamefont {R.}~\bibnamefont {Pain}},  \emph {et~al.},\ }\href {\doibase
  10.1051/0004-6361:20066930} {\bibfield  {journal} {\bibinfo  {journal}
  {Astron.Astrophys.}\ }\textbf {\bibinfo {volume} {466}},\ \bibinfo {pages}
  {11} (\bibinfo {year} {2007})},\ \Eprint
  {http://arxiv.org/abs/astro-ph/0701828} {astro-ph/0701828} \BibitemShut
  {NoStop}%
\bibitem [{\citenamefont {Eisenstein}\ \emph {et~al.}(2005)\citenamefont
  {Eisenstein} \emph {et~al.}}]{Eisenstein}%
  \BibitemOpen
  \bibfield  {author} {\bibinfo {author} {\bibfnamefont {D.~J.}\ \bibnamefont
  {Eisenstein}} \emph {et~al.},\ }\href@noop {} {\bibfield  {journal} {\bibinfo
   {journal} {Astrophys. J.}\ }\textbf {\bibinfo {volume} {633}},\ \bibinfo
  {pages} {560} (\bibinfo {year} {2005})},\ \Eprint
  {http://arxiv.org/abs/astro-ph/0501171} {astro-ph/0501171} \BibitemShut
  {NoStop}%
\bibitem [{\citenamefont {Beutler}\ \emph {et~al.}(2011)\citenamefont
  {Beutler}, \citenamefont {Blake}, \citenamefont {Colless}, \citenamefont
  {Jones}, \citenamefont {Staveley-Smith} \emph {et~al.}}]{Beutler:2011hx}%
  \BibitemOpen
  \bibfield  {author} {\bibinfo {author} {\bibfnamefont {F.}~\bibnamefont
  {Beutler}}, \bibinfo {author} {\bibfnamefont {C.}~\bibnamefont {Blake}},
  \bibinfo {author} {\bibfnamefont {M.}~\bibnamefont {Colless}}, \bibinfo
  {author} {\bibfnamefont {D.~H.}\ \bibnamefont {Jones}}, \bibinfo {author}
  {\bibfnamefont {L.}~\bibnamefont {Staveley-Smith}},  \emph {et~al.},\ }\href
  {\doibase 10.1111/j.1365-2966.2011.19250.x} {\bibfield  {journal} {\bibinfo
  {journal} {Mon.Not.Roy.Astron.Soc.}\ }\textbf {\bibinfo {volume} {416}},\
  \bibinfo {pages} {3017} (\bibinfo {year} {2011})},\ \Eprint
  {http://arxiv.org/abs/1106.3366} {arXiv:1106.3366} \BibitemShut {NoStop}%
\bibitem [{\citenamefont {Percival}\ \emph {et~al.}(2010)\citenamefont
  {Percival} \emph {et~al.}}]{Percival:2009xn}%
  \BibitemOpen
  \bibfield  {author} {\bibinfo {author} {\bibfnamefont {W.~J.}\ \bibnamefont
  {Percival}} \emph {et~al.} (\bibinfo {collaboration} {SDSS Collaboration}),\
  }\href {\doibase 10.1111/j.1365-2966.2009.15812.x} {\bibfield  {journal}
  {\bibinfo  {journal} {Mon.Not.Roy.Astron.Soc.}\ }\textbf {\bibinfo {volume}
  {401}},\ \bibinfo {pages} {2148} (\bibinfo {year} {2010})},\ \Eprint
  {http://arxiv.org/abs/0907.1660} {arXiv:0907.1660} \BibitemShut {NoStop}%
\bibitem [{\citenamefont {Blake}\ \emph {et~al.}(2012)\citenamefont {Blake},
  \citenamefont {Brough}, \citenamefont {Colless}, \citenamefont {Contreras},
  \citenamefont {Couch} \emph {et~al.}}]{Blake:2012pj}%
  \BibitemOpen
  \bibfield  {author} {\bibinfo {author} {\bibfnamefont {C.}~\bibnamefont
  {Blake}}, \bibinfo {author} {\bibfnamefont {S.}~\bibnamefont {Brough}},
  \bibinfo {author} {\bibfnamefont {M.}~\bibnamefont {Colless}}, \bibinfo
  {author} {\bibfnamefont {C.}~\bibnamefont {Contreras}}, \bibinfo {author}
  {\bibfnamefont {W.}~\bibnamefont {Couch}},  \emph {et~al.},\ }\href@noop {}
  {\  (\bibinfo {year} {2012})},\ \Eprint {http://arxiv.org/abs/1204.3674}
  {arXiv:1204.3674} \BibitemShut {NoStop}%
\bibitem [{\citenamefont {Blake}\ \emph {et~al.}(2011)\citenamefont {Blake},
  \citenamefont {Kazin}, \citenamefont {Beutler}, \citenamefont {Davis},
  \citenamefont {Parkinson} \emph {et~al.}}]{Blake:2011en}%
  \BibitemOpen
  \bibfield  {author} {\bibinfo {author} {\bibfnamefont {C.}~\bibnamefont
  {Blake}}, \bibinfo {author} {\bibfnamefont {E.}~\bibnamefont {Kazin}},
  \bibinfo {author} {\bibfnamefont {F.}~\bibnamefont {Beutler}}, \bibinfo
  {author} {\bibfnamefont {T.}~\bibnamefont {Davis}}, \bibinfo {author}
  {\bibfnamefont {D.}~\bibnamefont {Parkinson}},  \emph {et~al.},\ }\href
  {\doibase 10.1111/j.1365-2966.2011.19592.x} {\bibfield  {journal} {\bibinfo
  {journal} {Mon.Not.Roy.Astron.Soc.}\ }\textbf {\bibinfo {volume} {418}},\
  \bibinfo {pages} {1707} (\bibinfo {year} {2011})},\ \Eprint
  {http://arxiv.org/abs/1108.2635} {arXiv:1108.2635} \BibitemShut {NoStop}%
\bibitem [{\citenamefont {Sanchez}\ \emph {et~al.}(2012)\citenamefont
  {Sanchez}, \citenamefont {Scoccola}, \citenamefont {Ross}, \citenamefont
  {Percival}, \citenamefont {Manera} \emph {et~al.}}]{Sanchez:2012sg}%
  \BibitemOpen
  \bibfield  {author} {\bibinfo {author} {\bibfnamefont {A.~G.}\ \bibnamefont
  {Sanchez}}, \bibinfo {author} {\bibfnamefont {C.}~\bibnamefont {Scoccola}},
  \bibinfo {author} {\bibfnamefont {A.}~\bibnamefont {Ross}}, \bibinfo {author}
  {\bibfnamefont {W.}~\bibnamefont {Percival}}, \bibinfo {author}
  {\bibfnamefont {M.}~\bibnamefont {Manera}},  \emph {et~al.},\ }\href@noop {}
  {\  (\bibinfo {year} {2012})},\ \Eprint {http://arxiv.org/abs/1203.6616}
  {arXiv:1203.6616} \BibitemShut {NoStop}%
\bibitem [{\citenamefont {Anderson}\ \emph {et~al.}(2012)\citenamefont
  {Anderson}, \citenamefont {Aubourg}, \citenamefont {Bailey}, \citenamefont
  {Bizyaev}, \citenamefont {Blanton} \emph {et~al.}}]{Anderson:2012sa}%
  \BibitemOpen
  \bibfield  {author} {\bibinfo {author} {\bibfnamefont {L.}~\bibnamefont
  {Anderson}}, \bibinfo {author} {\bibfnamefont {E.}~\bibnamefont {Aubourg}},
  \bibinfo {author} {\bibfnamefont {S.}~\bibnamefont {Bailey}}, \bibinfo
  {author} {\bibfnamefont {D.}~\bibnamefont {Bizyaev}}, \bibinfo {author}
  {\bibfnamefont {M.}~\bibnamefont {Blanton}},  \emph {et~al.},\ }\href@noop {}
  {\  (\bibinfo {year} {2012})},\ \Eprint {http://arxiv.org/abs/1203.6594}
  {arXiv:1203.6594} \BibitemShut {NoStop}%
\bibitem [{\citenamefont {Frieman}\ \emph {et~al.}(2003)\citenamefont
  {Frieman}, \citenamefont {Huterer}, \citenamefont {Linder},\ and\
  \citenamefont {Turner}}]{FriHutLinTur}%
  \BibitemOpen
  \bibfield  {author} {\bibinfo {author} {\bibfnamefont {J.~A.}\ \bibnamefont
  {Frieman}}, \bibinfo {author} {\bibfnamefont {D.}~\bibnamefont {Huterer}},
  \bibinfo {author} {\bibfnamefont {E.~V.}\ \bibnamefont {Linder}}, \ and\
  \bibinfo {author} {\bibfnamefont {M.~S.}\ \bibnamefont {Turner}},\
  }\href@noop {} {\bibfield  {journal} {\bibinfo  {journal} {Phys. Rev.}\
  }\textbf {\bibinfo {volume} {D67}},\ \bibinfo {pages} {083505} (\bibinfo
  {year} {2003})},\ \Eprint {http://arxiv.org/abs/astro-ph/0208100}
  {astro-ph/0208100} \BibitemShut {NoStop}%
\bibitem [{\citenamefont {Komatsu}\ \emph {et~al.}(2011)\citenamefont {Komatsu}
  \emph {et~al.}}]{wmap7}%
  \BibitemOpen
  \bibfield  {author} {\bibinfo {author} {\bibfnamefont {E.}~\bibnamefont
  {Komatsu}} \emph {et~al.} (\bibinfo {collaboration} {WMAP}),\ }\href@noop {}
  {\bibfield  {journal} {\bibinfo  {journal} {Astrophys. J. Suppl.}\ }\textbf
  {\bibinfo {volume} {192}},\ \bibinfo {pages} {18} (\bibinfo {year} {2011})},\
  \Eprint {http://arxiv.org/abs/1001.4538} {arXiv:1001.4538} \BibitemShut
  {NoStop}%
\bibitem [{\citenamefont {Christensen}\ \emph {et~al.}(2001)\citenamefont
  {Christensen}, \citenamefont {Meyer}, \citenamefont {Knox},\ and\
  \citenamefont {Luey}}]{Christensen:2001gj}%
  \BibitemOpen
  \bibfield  {author} {\bibinfo {author} {\bibfnamefont {N.}~\bibnamefont
  {Christensen}}, \bibinfo {author} {\bibfnamefont {R.}~\bibnamefont {Meyer}},
  \bibinfo {author} {\bibfnamefont {L.}~\bibnamefont {Knox}}, \ and\ \bibinfo
  {author} {\bibfnamefont {B.}~\bibnamefont {Luey}},\ }\href@noop {} {\bibfield
   {journal} {\bibinfo  {journal} {Class. Quant. Grav.}\ }\textbf {\bibinfo
  {volume} {18}},\ \bibinfo {pages} {2677} (\bibinfo {year} {2001})},\ \Eprint
  {http://arxiv.org/abs/astro-ph/0103134} {astro-ph/0103134} \BibitemShut
  {NoStop}%
\bibitem [{\citenamefont {{Dunkley}}\ \emph {et~al.}(2005)\citenamefont
  {{Dunkley}}, \citenamefont {{Bucher}}, \citenamefont {{Ferreira}},
  \citenamefont {{Moodley}},\ and\ \citenamefont {{Skordis}}}]{Dunetal05}%
  \BibitemOpen
  \bibfield  {author} {\bibinfo {author} {\bibfnamefont {J.}~\bibnamefont
  {{Dunkley}}}, \bibinfo {author} {\bibfnamefont {M.}~\bibnamefont {{Bucher}}},
  \bibinfo {author} {\bibfnamefont {P.~G.}\ \bibnamefont {{Ferreira}}},
  \bibinfo {author} {\bibfnamefont {K.}~\bibnamefont {{Moodley}}}, \ and\
  \bibinfo {author} {\bibfnamefont {C.}~\bibnamefont {{Skordis}}},\ }\href
  {\doibase 10.1111/j.1365-2966.2004.08464.x} {\bibfield  {journal} {\bibinfo
  {journal} {\mnras}\ }\textbf {\bibinfo {volume} {356}},\ \bibinfo {pages}
  {925} (\bibinfo {year} {2005})},\ \Eprint
  {http://arxiv.org/abs/astro-ph/0405462} {astro-ph/0405462} \BibitemShut
  {NoStop}%
\bibitem [{\citenamefont {Metropolis}\ \emph {et~al.}(1953)\citenamefont
  {Metropolis}, \citenamefont {Rosenbluth}, \citenamefont {Rosenbluth},
  \citenamefont {Teller},\ and\ \citenamefont {Teller}}]{Metropolis:1953am}%
  \BibitemOpen
  \bibfield  {author} {\bibinfo {author} {\bibfnamefont {N.}~\bibnamefont
  {Metropolis}}, \bibinfo {author} {\bibfnamefont {A.}~\bibnamefont
  {Rosenbluth}}, \bibinfo {author} {\bibfnamefont {M.}~\bibnamefont
  {Rosenbluth}}, \bibinfo {author} {\bibfnamefont {A.}~\bibnamefont {Teller}},
  \ and\ \bibinfo {author} {\bibfnamefont {E.}~\bibnamefont {Teller}},\ }\href
  {\doibase 10.1063/1.1699114} {\bibfield  {journal} {\bibinfo  {journal}
  {J.Chem.Phys.}\ }\textbf {\bibinfo {volume} {21}},\ \bibinfo {pages} {1087}
  (\bibinfo {year} {1953})}\BibitemShut {NoStop}%
\bibitem [{\citenamefont {Hastings}(1970)}]{hastings}%
  \BibitemOpen
  \bibfield  {author} {\bibinfo {author} {\bibfnamefont {W.}~\bibnamefont
  {Hastings}},\ }\href {\doibase 10.1093/biomet/57.1.97} {\bibfield  {journal}
  {\bibinfo  {journal} {Biometrika}\ }\textbf {\bibinfo {volume} {57}},\
  \bibinfo {pages} {97} (\bibinfo {year} {1970})}\BibitemShut {NoStop}%
\bibitem [{\citenamefont {Gelman}\ and\ \citenamefont
  {Rubin}(1992)}]{gelman/rubin}%
  \BibitemOpen
  \bibfield  {author} {\bibinfo {author} {\bibfnamefont {A.}~\bibnamefont
  {Gelman}}\ and\ \bibinfo {author} {\bibfnamefont {D.}~\bibnamefont {Rubin}},\
  }\href@noop {} {\bibfield  {journal} {\bibinfo  {journal} {Statistical
  Science}\ }\textbf {\bibinfo {volume} {7}},\ \bibinfo {pages} {452} (\bibinfo
  {year} {1992})}\BibitemShut {NoStop}%
\bibitem [{\citenamefont {Lewis}\ and\ \citenamefont
  {Bridle}(2002)}]{Lewis:2002ah}%
  \BibitemOpen
  \bibfield  {author} {\bibinfo {author} {\bibfnamefont {A.}~\bibnamefont
  {Lewis}}\ and\ \bibinfo {author} {\bibfnamefont {S.}~\bibnamefont {Bridle}},\
  }\href@noop {} {\bibfield  {journal} {\bibinfo  {journal} {Phys. Rev.}\
  }\textbf {\bibinfo {volume} {D66}},\ \bibinfo {pages} {103511} (\bibinfo
  {year} {2002})},\ \Eprint {http://arxiv.org/abs/astro-ph/0205436}
  {astro-ph/0205436} \BibitemShut {NoStop}%
\bibitem [{cos()}]{cosmomc_url}%
  \BibitemOpen
  \href@noop {} {\ }\bibinfo {note}
  {\url{http://cosmologist.info/cosmomc/}}\BibitemShut {NoStop}%
\bibitem [{\citenamefont {Linder}(2003)}]{Linder_wa}%
  \BibitemOpen
  \bibfield  {author} {\bibinfo {author} {\bibfnamefont {E.~V.}\ \bibnamefont
  {Linder}},\ }\href@noop {} {\bibfield  {journal} {\bibinfo  {journal} {Phys.
  Rev. Lett.}\ }\textbf {\bibinfo {volume} {90}},\ \bibinfo {pages} {091301}
  (\bibinfo {year} {2003})},\ \Eprint {http://arxiv.org/abs/astro-ph/0208512}
  {astro-ph/0208512} \BibitemShut {NoStop}%
\bibitem [{\citenamefont {Huterer}\ and\ \citenamefont
  {Starkman}(2003)}]{Huterer_Starkman}%
  \BibitemOpen
  \bibfield  {author} {\bibinfo {author} {\bibfnamefont {D.}~\bibnamefont
  {Huterer}}\ and\ \bibinfo {author} {\bibfnamefont {G.}~\bibnamefont
  {Starkman}},\ }\href@noop {} {\bibfield  {journal} {\bibinfo  {journal}
  {Phys. Rev. Lett.}\ }\textbf {\bibinfo {volume} {90}},\ \bibinfo {pages}
  {031301} (\bibinfo {year} {2003})},\ \Eprint
  {http://arxiv.org/abs/astro-ph/0207517} {astro-ph/0207517} \BibitemShut
  {NoStop}%
\bibitem [{\citenamefont {Chevallier}\ and\ \citenamefont
  {Polarski}(2001)}]{Chevallier_Polarski}%
  \BibitemOpen
  \bibfield  {author} {\bibinfo {author} {\bibfnamefont {M.}~\bibnamefont
  {Chevallier}}\ and\ \bibinfo {author} {\bibfnamefont {D.}~\bibnamefont
  {Polarski}},\ }\href@noop {} {\bibfield  {journal} {\bibinfo  {journal} {Int.
  J. Mod. Phys.}\ }\textbf {\bibinfo {volume} {D10}},\ \bibinfo {pages} {213}
  (\bibinfo {year} {2001})},\ \Eprint {http://arxiv.org/abs/gr-qc/0009008}
  {gr-qc/0009008} \BibitemShut {NoStop}%
\bibitem [{\citenamefont {Albrecht}\ \emph {et~al.}(2006)\citenamefont
  {Albrecht} \emph {et~al.}}]{DETF}%
  \BibitemOpen
  \bibfield  {author} {\bibinfo {author} {\bibfnamefont {A.}~\bibnamefont
  {Albrecht}} \emph {et~al.},\ }\href@noop {} {\  (\bibinfo {year} {2006})},\
  \Eprint {http://arxiv.org/abs/astro-ph/0609591} {astro-ph/0609591}
  \BibitemShut {NoStop}%
\bibitem [{\citenamefont {Huterer}\ and\ \citenamefont
  {Turner}(2001)}]{Huterer_Turner}%
  \BibitemOpen
  \bibfield  {author} {\bibinfo {author} {\bibfnamefont {D.}~\bibnamefont
  {Huterer}}\ and\ \bibinfo {author} {\bibfnamefont {M.~S.}\ \bibnamefont
  {Turner}},\ }\href@noop {} {\bibfield  {journal} {\bibinfo  {journal} {Phys.
  Rev.}\ }\textbf {\bibinfo {volume} {D64}},\ \bibinfo {pages} {123527}
  (\bibinfo {year} {2001})},\ \Eprint {http://arxiv.org/abs/astro-ph/0012510}
  {astro-ph/0012510} \BibitemShut {NoStop}%
\bibitem [{\citenamefont {Albrecht}\ \emph {et~al.}(2009)\citenamefont
  {Albrecht} \emph {et~al.}}]{FoMSWG}%
  \BibitemOpen
  \bibfield  {author} {\bibinfo {author} {\bibfnamefont {A.~J.}\ \bibnamefont
  {Albrecht}} \emph {et~al.},\ }\href@noop {} {\  (\bibinfo {year} {2009})},\
  \Eprint {http://arxiv.org/abs/0901.0721} {arXiv:0901.0721} \BibitemShut
  {NoStop}%
\bibitem [{\citenamefont {{Mortonson}}\ \emph {et~al.}(2009)\citenamefont
  {{Mortonson}}, \citenamefont {{Hu}},\ and\ \citenamefont
  {{Huterer}}}]{Mort_falsifying}%
  \BibitemOpen
  \bibfield  {author} {\bibinfo {author} {\bibfnamefont {M.~J.}\ \bibnamefont
  {{Mortonson}}}, \bibinfo {author} {\bibfnamefont {W.}~\bibnamefont {{Hu}}}, \
  and\ \bibinfo {author} {\bibfnamefont {D.}~\bibnamefont {{Huterer}}},\ }\href
  {\doibase 10.1103/PhysRevD.79.023004} {\bibfield  {journal} {\bibinfo
  {journal} {Phys. Rev.}\ }\textbf {\bibinfo {volume} {D79}},\ \bibinfo {pages}
  {023004} (\bibinfo {year} {2009})},\ \Eprint {http://arxiv.org/abs/0810.1744}
  {arXiv:0810.1744} \BibitemShut {NoStop}%
\bibitem [{\citenamefont {Bassett}\ and\ \citenamefont
  {Afshordi}(2010)}]{Bassett_Afshordi}%
  \BibitemOpen
  \bibfield  {author} {\bibinfo {author} {\bibfnamefont {B.~A.}\ \bibnamefont
  {Bassett}}\ and\ \bibinfo {author} {\bibfnamefont {N.}~\bibnamefont
  {Afshordi}},\ }\href@noop {} {\  (\bibinfo {year} {2010})},\ \Eprint
  {http://arxiv.org/abs/1005.1664} {arXiv:1005.1664} \BibitemShut {NoStop}%
\bibitem [{\citenamefont {Kessler}\ \emph {et~al.}(2010)\citenamefont
  {Kessler}, \citenamefont {Bassett}, \citenamefont {Belov}, \citenamefont
  {Bhatnagar}, \citenamefont {Campbell} \emph
  {et~al.}}]{Kessler_photo_classification}%
  \BibitemOpen
  \bibfield  {author} {\bibinfo {author} {\bibfnamefont {R.}~\bibnamefont
  {Kessler}}, \bibinfo {author} {\bibfnamefont {B.}~\bibnamefont {Bassett}},
  \bibinfo {author} {\bibfnamefont {P.}~\bibnamefont {Belov}}, \bibinfo
  {author} {\bibfnamefont {V.}~\bibnamefont {Bhatnagar}}, \bibinfo {author}
  {\bibfnamefont {H.}~\bibnamefont {Campbell}},  \emph {et~al.},\ }\href
  {\doibase 10.1086/657607} {\bibfield  {journal} {\bibinfo  {journal}
  {Publ.Astron.Soc.Pac.}\ }\textbf {\bibinfo {volume} {122}},\ \bibinfo {pages}
  {1415} (\bibinfo {year} {2010})},\ \Eprint {http://arxiv.org/abs/1008.1024}
  {arXiv:1008.1024} \BibitemShut {NoStop}%
\bibitem [{\citenamefont {Hlozek}\ \emph {et~al.}(2012)\citenamefont {Hlozek},
  \citenamefont {Kunz}, \citenamefont {Bassett}, \citenamefont {Smith},
  \citenamefont {Newling} \emph {et~al.}}]{BEAMS}%
  \BibitemOpen
  \bibfield  {author} {\bibinfo {author} {\bibfnamefont {R.}~\bibnamefont
  {Hlozek}}, \bibinfo {author} {\bibfnamefont {M.}~\bibnamefont {Kunz}},
  \bibinfo {author} {\bibfnamefont {B.}~\bibnamefont {Bassett}}, \bibinfo
  {author} {\bibfnamefont {M.}~\bibnamefont {Smith}}, \bibinfo {author}
  {\bibfnamefont {J.}~\bibnamefont {Newling}},  \emph {et~al.},\ }\href
  {\doibase 10.1088/0004-637X/752/2/79} {\bibfield  {journal} {\bibinfo
  {journal} {Astrophys.J.}\ }\textbf {\bibinfo {volume} {752}},\ \bibinfo
  {pages} {79} (\bibinfo {year} {2012})},\ \Eprint
  {http://arxiv.org/abs/1111.5328} {arXiv:1111.5328} \BibitemShut {NoStop}%
\end{thebibliography}%

\end{document}